\documentclass[letter,iop]{emulateapj}
\usepackage{amssymb}
\usepackage{natbib}
\usepackage{graphicx}

\def\N1Mpc{$N_{\rm 1Mpc}$\/ }
\def\L200{$L_{200}$\/}
\def\n200{$N_{200}^{\rm gal}$\/}
\def\LBCG{$L_{\rm BCG}$\/}
\def\MN200{$M(N_{200})$\/}
\def\ML200{$M(L_{200})$\/}

\def\ltsima{$\; \buildrel < \over \sim \;$}
\def\simlt{\lower.5ex\hbox{\ltsima}}
\def\gtsima{$\; \buildrel > \over \sim \;$}
\def\simgt{\lower.5ex\hbox{\gtsima}}
\def\simless{\mathbin{\lower 3pt\hbox
   {$\rlap{\raise 5pt\hbox{$\char'074$}}\mathchar"7218$}}}   % < or of order
\def\simgreat{\mathbin{\lower 3pt\hbox
   {$\rlap{\raise 5pt\hbox{$\char'076$}}\mathchar"7218$}}}   % > or of order

\shorttitle{Southern Cosmology Survey II}
\shortauthors{Menanteau et al.}
\submitted{ApJSS accepted}

\begin{document}

\title{Southern Cosmology Survey II: Massive Optically-Selected Clusters from
  70 square degrees of the SZE Common Survey Area}
\author{Felipe Menanteau\altaffilmark{1},
John P. Hughes\altaffilmark{1},
L. Felipe Barrientos\altaffilmark{2},
Amruta J. Deshpande\altaffilmark{1},
Matt Hilton\altaffilmark{3},
Leopoldo~Infante\altaffilmark{2},
Raul Jimenez\altaffilmark{4}, 
Arthur Kosowsky\altaffilmark{5},
Kavilan Moodley\altaffilmark{3},
David Spergel\altaffilmark{6}, and
Licia Verde\altaffilmark{4}
}

% ms mode 
%\altaffiltext{1}{Rutgers University, Department of Physics \& Astronomy, 136 Frelinghuysen Rd, Piscataway, NJ 08854-8019, USA }
%\altaffiltext{2}{Pontificia Universidad Cat\'olica de Chile, Departamento de Astronom\'{i}a, Santiago, Chile}
%\altaffiltext{3}{University of KwaZulu-Natal, Astrophysics \& Cosmology Research Unit, School of Mathematical Sciences, Durban, 4041, South Africa.}
%\altaffiltext{4}{ICREA \& Institute for Sciences of the Cosmos (ICC), University of Barcelona, Marti i Franques 1, Barcelona 08034, Spain}
%\altaffiltext{5}{University of Pittsburgh, Physics \& Astronomy Department, 100 Allen Hall, 3941 O'Hara Street, Pittsburgh, PA 15260, USA}
%\altaffiltext{6}{Department of Astrophysical Sciences, Peyton Hall, Princeton University, Princeton, NJ 08544, USA}

% emulateapj mode
\affil{$^1$Rutgers University, Department of Physics \& Astronomy, 136 Frelinghuysen Rd, Piscataway, NJ 08854, USA }
\affil{$^2$Pontificia Universidad Cat\'olica de Chile, Departamento de Astronom\'{i}a, Santiago, Chile}
\affil{$^3$University of KwaZulu-Natal, Astrophysics \& Cosmology Research Unit, School of Mathematical Sciences, Durban, 4041, South Africa.}
\affil{$^4$ICREA \& Institute for Sciences of the Cosmos (ICC), University of Barcelona, Marti i Franques 1, Barcelona 08034, Spain}
\affil{$^5$University of Pittsburgh, Physics \& Astronomy Department, 100 Allen Hall, 3941 O'Hara Street, Pittsburgh, PA 15260, USA}
\affil{$^6$Department of Astrophysical Sciences, Peyton Hall, Princeton University, Princeton, NJ 08544, USA}

\begin{abstract}

We present a catalog of 105 rich and massive
($M>3\times10^{14}M_{\sun}$) optically-selected clusters of galaxies
extracted from 70 square-degrees of public archival $griz$ imaging
from the Blanco 4-m telescope acquired over 45 nights between 2005 and
2007.  We use the clusters' optically-derived properties to estimate
photometric redshifts, optical luminosities, richness, and masses. We
complement the optical measurements with archival XMM-Newton and ROSAT
X-ray data which provide additional luminosity and mass constraints on
a modest fraction of the cluster sample. Two of our
clusters show clear evidence for central lensing arcs; one of these has
a spectacular large-diameter, nearly-complete
Einstein Ring surrounding the brightest cluster galaxy.
A strong motivation for this study is to identify the
massive clusters that are expected to display prominent signals from
the Sunyaev-Zeldovich Effect (SZE) and therefore be detected in the
wide-area mm-band surveys being conducted by both the Atacama
Cosmology Telescope and the South Pole Telescope. The optical sample
presented here will be useful for verifying new SZE cluster candidates
from these surveys, for testing the cluster selection function, and
for stacking analyzes of the SZE data.

\end{abstract}


\section{Introduction}

A new era of galaxy cluster surveys, based on measuring distortions in
the cosmic microwave background (CMB), has begun.  These distortions,
known as the Sunyaev-Zel'dovich Effect \citep{SZ72}, have been
detected for the first time in untargeted surveys over large areas of
the sky by two new mm-band experiments: the Atacama Cosmology
Telescope (ACT) and the South Pole Telescope (SPT). First results from
ACT \citep{Hincks09} and SPT \citep{Stan09} offer a taste of the
future potential these experiments hold for obtaining large samples of
essentially mass-selected clusters to arbitrary redshifts and have
also provided the first measurement of the microwave background at
arc-minute angular scales \citep{SPT-CMB, ACT-CMB}.

Both ACT and SPT aim to provide unique samples of massive clusters of
galaxies, selected by mass nearly independently of redshift, over a
large area of the southern sky. While the number density of
SZE-selected clusters can be used as a potentially strong probe of
dark energy---as well as for studies of cluster physics, it is crucial
to understand the systematics of SZE surveys by comparing with cluster
identification using independent methods before the new cluster
samples can be effectively used. 
For example the low amplitude of the SZ component in the high-$l$ CMB
power spectrum \citep{SPT-CMB} or stacked clusters in WMAP
\citep{Komatsu10} are recent issues that provide additional motivation
for an independent search for clusters over the region being surveyed
in the SZ.  Furthermore, although the new mm-band telescopes can be
used to identify clusters, coordinated optical data are necessary for
confirmation and to determine redshifts and other fundamental
properties of the new clusters.

The last decade has seen significant effort to produce large and
well-selected optical catalogs of cluster of galaxies that can be used
in cosmological, large-scale structure and galaxy evolution
studies. The first systematic attempts to generate large samples of
clusters, and to define their richness, came from the Abell catalogs
\citep{Abell-58,ACO89} which searched for projected galaxy overdensities
through visual inspection of photographic plates successfully
identifying thousands of clusters.
Although optical catalogs can be relatively inexpensive and efficient
at detecting low mass systems, early attempts were known to suffer
from significant projection effects along the line of sight.
The advent of CCD cameras and the digitization of large photographic
plates has enabled the development of new search algorithms for galaxy
clusters using a combination of space, brightness and color
information (i.e. photometric redshifts), minimizing projection issues
\citep[see][for a historical review of search
methods]{Gal-review}. Among these algorithms are the pioneering
implementation of a spatial matched filter technique \citep{Postman96}
and its variants \citep{Dong08}, the adaptive kernel technique
\citep{NoSOCSI,NoSOCSIII}, Voronoi tessellation
\citep{Ramella01,Lopes04} and methods exploiting the tight ridgeline
in color-magnitude space of galaxies in clusters
\citep{Bower92,Blakeslee03} such as the Red Cluster Sequence (RCS)
\citep{Gladders-Yee} and the MaxBCG \citep{Annis99,MaxBCG}
techniques. This new wave of studies have produced large sets of
well-defined optical cluster catalogs covering thousands of
square-degrees \citep[i.e.][]{Goto02,RCS,MaxBCG-Cat,NoSOCSIII}
providing reliable richness-mass correlations
\citep{Becker07,Johnston07,Reyes08,Sheldon09} and establishing
independent cosmological constraints \citep{Rozo10} using optical
catalogs.

In this article we present new results from the Southern Cosmology
Survey (SCS), our on-going multi-wavelength survey coordinated with
ACT. Our first paper in this series, \citet{SCSI}, described our SCS 
imaging pipeline and presented a sample of new galaxy
clusters from an $8$~deg$^2$ optical
imaging survey of the southern sky acquired at the Blanco 4-m
telescope. Here, we complete our cluster analysis using all the
70~deg$^2$ contiguous imaging available, which represents a
comoving volume of 0.076~Gpc$^3$ at $z=0.6$.
% ---- *** ------
Throughout this paper we assume a flat cosmology with $H_0=100
h$~km~s$^{-1}$~Mpc$^{-1}$, $h=0.7$ and matter density $\Omega_M=0.3$.

\section{Data Set and Analysis}

\begin{figure}
\centerline{\includegraphics[width=3.5in]{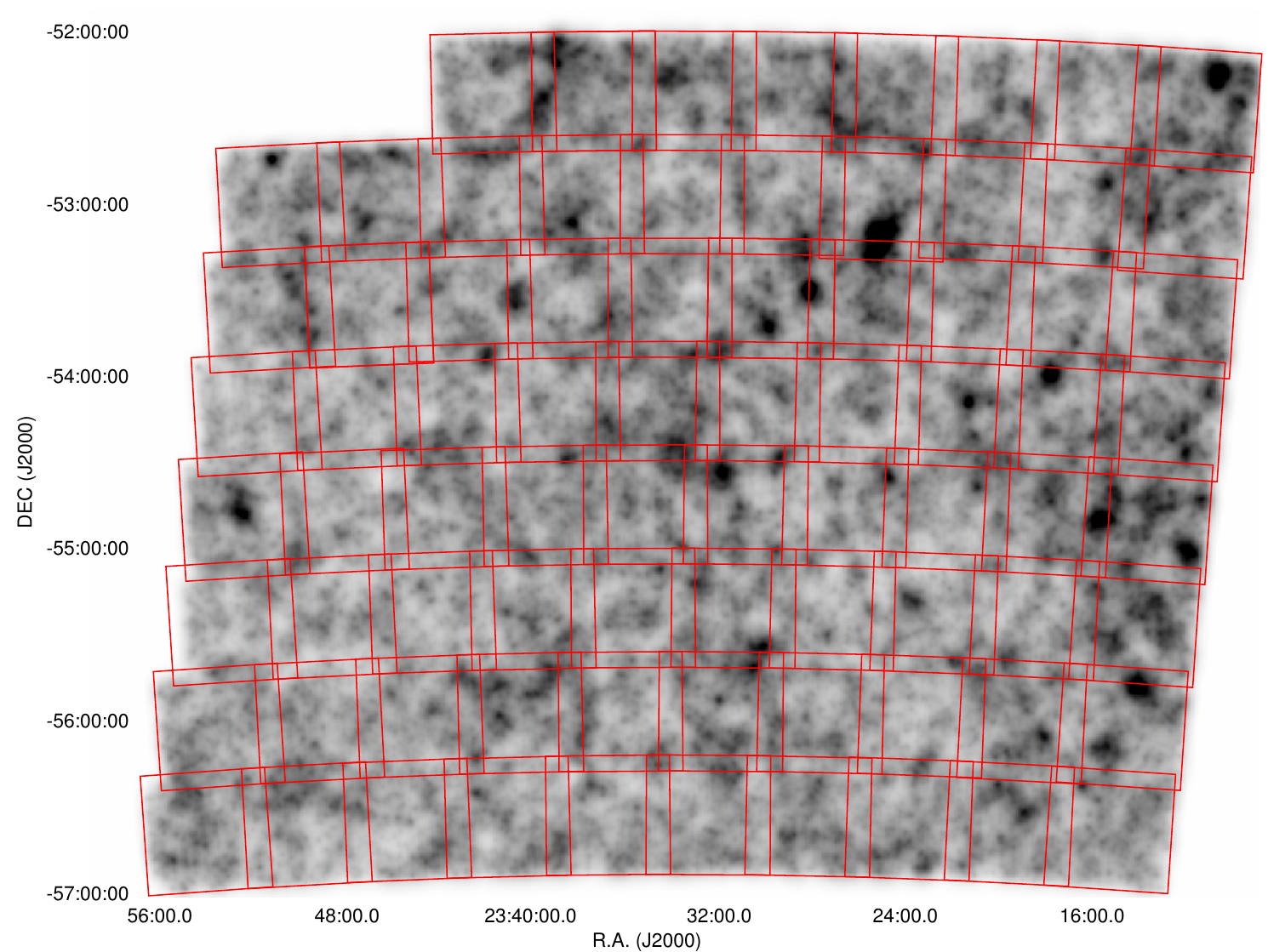}}
%\centerline{\includegraphics[width=3.9in]{f1a.pdf}}
\vspace{0.5cm}
\centerline{\includegraphics[width=3.6in]{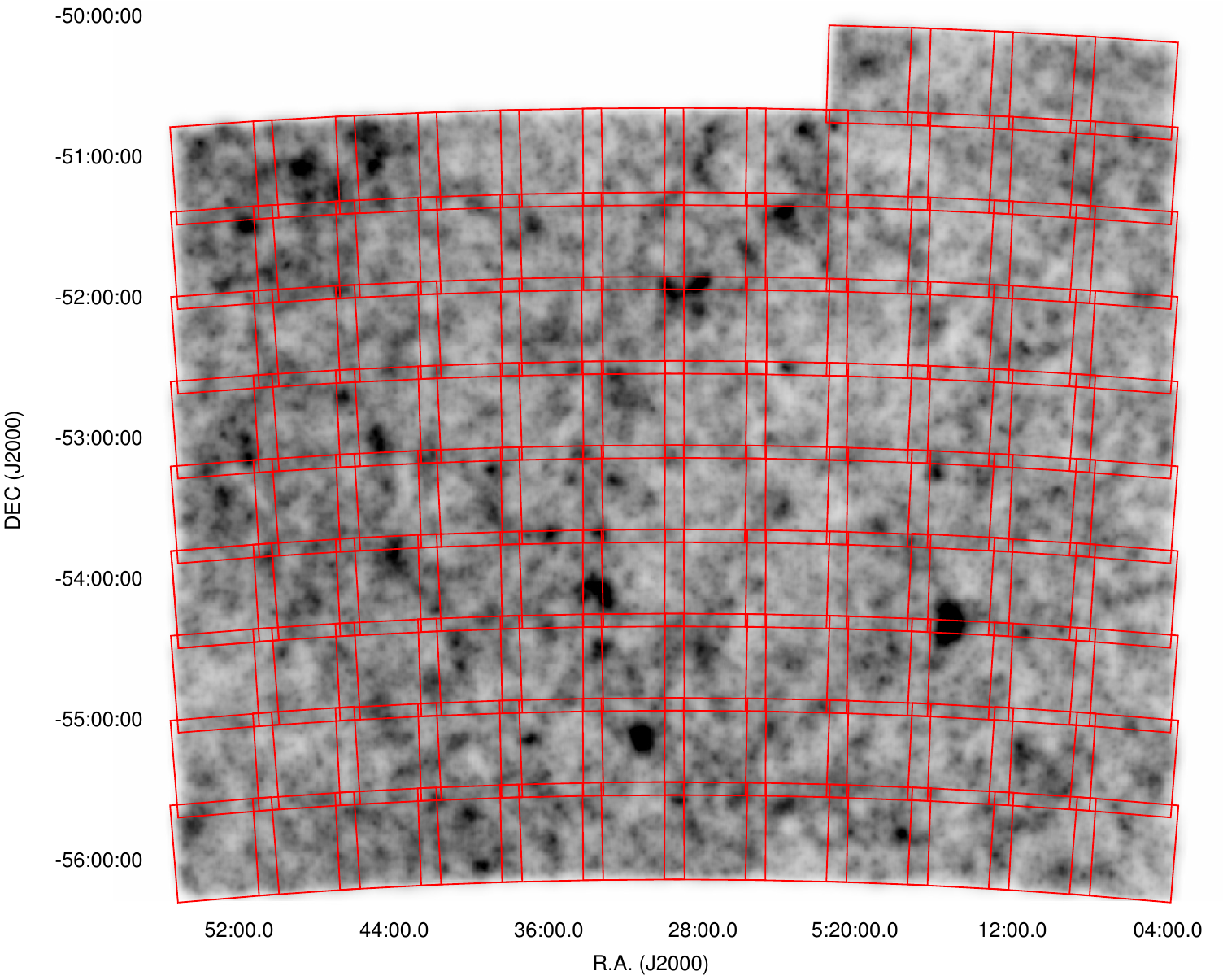}}
%\centerline{\includegraphics[width=4.2in]{f1b.pdf}}
\caption{The computed likelihood density map images centered at
  $z=0.3$ and width $\Delta z=0.1$ over the 23hr (top panel) 5hr
  (bottom panel) field. Dark regions in the image represent denser
  areas. In red we show the area covered by each of the 78 and 112
  MOSAIC-II tiles at 23hr and 5hr respectively that comprise the area
  studied over three years of observations.}
\label{fig:densemap}
\end{figure}

Our cluster analysis is based on all of the public unprocessed imaging
data available for download at the time of writing from the
observations carried out for the Blanco Cosmology
Survey\footnote{http://cosmology.uiuc.edu/BCS/} (BCS) proposal. This
was a NOAO Large Survey project (05B-0043, PI: Joe Mohr)
that was awarded 45 nights between 2005 and 2007 to carry
out intermediate depth $g,r,i,z$--band observations on the Cerro
Tololo InterAmerican Observatory (CTIO) Blanco 4-m telescope using the
$8192\times8192$ pixel (0.36~deg$^2$) MOSAIC-II camera.  The survey
originally aimed to target two square $50$~deg$^2$ southern sky
patches centered near declinations $-55^{\circ}$ and $-52^{\circ}$ at
right ascensions of 23 hr and 5 hr respectively, that were to be
contained in a larger SZE common region that both ACT and SPT would
survey. After three years of observations the survey acquired 78 and
118 MOSAIC-II contiguous pointings at 23hr and 5hr respectively that
correspond to 29~deg$^2$ and 41~deg$^2$ over each patch reaching a
total of 70~deg$^2$ imaged in all $griz$ filters. The pointing layout
and the sky area covered for each patch is shown in
Figure~\ref{fig:densemap}. The survey was awarded 15 extra nights at
the end of 2008 to complete the proposed 100~deg$^2$; our analysis
does not include this extra year's worth of observations as the data
are not publicly available yet.

\subsection{Data Processing}

The raw unprocessed images were downloaded from the National Virtual
Observatory\footnote{http://portal-nvo.noao.edu} portal and were
processed following the same procedures as described in our initial
analysis \citep{SCSI} where we provide a full description of the
data analysis followed and associated data products.
Here we briefly outline the steps involved and recent improvements
from our original analysis. Our pipeline automatically handles the
file associations and all of the initial standard CCD imaging tasks
for each night's run as well as the secondary calibration steps (i.e.,
skyflats, fringe correction, cosmic-ray rejection, badpixel masks,
world coordinate calibration, etc.) using a custom modified version of
IRAF\footnote{IRAF is distributed by the National Optical Astronomy
  Observatory, which is operated by the AURA under cooperative
  agreement with NSF}/{\tt mscred} \citep{Valdes98}. When observed, photometric
standards from the Southern Hemisphere Standards Stars Catalog
\citep{Smith07} were processed by the pipeline and a photometric
zero-point was estimated in AB magnitudes. Hereafter all magnitude
quoted are in the AB system. In table~\ref{tab:obs} we show the
observing dates, photometric conditions and observed bands that
comprise the dataset we analyzed and that were used for this paper. We
note that the information that went into creating this table was
extracted from the raw data and header information of the data files.

\begin{figure*}
%\centerline{\includegraphics[width=6.2in]{f2.pdf}}
\centerline{\includegraphics[width=6in]{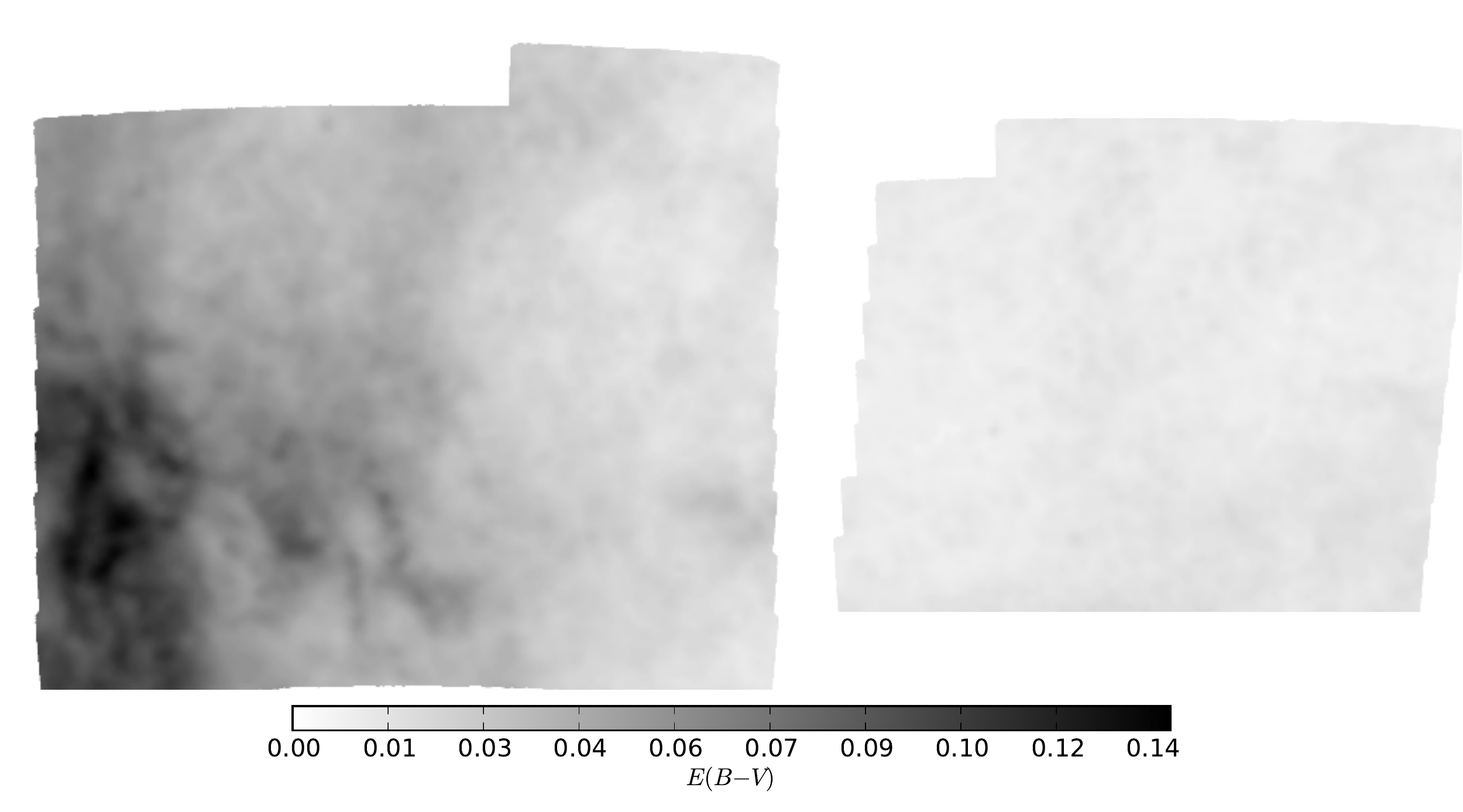}}
\caption{The $E(B-V)$ dust extinction maps created for the 5hr (left)
  and 23hr (right) fields in magnitudes using the \cite{Schlegel98}
  maps of dust infrared emission. Both maps are showed with the same
  $E(B-V)$ scale and size scales. Darker regions represent areas of
  higher Galactic dust absorption.}
\label{fig:dustmap}
\end{figure*}

In general each MOSAIC~II pointing consisted of exposures of
$2\times125s$, $2\times300s$, $3\times450s$ and $3\times235s$ in the
$g,r,i$ and $z-$bands respectively with offsets of $3-5$~arcmin
(within each filter). We adjusted the photometric zero points of
non-photometric nights using the overlapping regions between tiles and
their matched sources from adjacent photometric tiles providing a
homogeneous photometric calibration across each region with typical
variations below $0.03$~mags. The 2007 season data at 23hr was
particularly hard to match due to poor observing conditions and large
variations within the individual exposures in each tile. We tackled
this problem by first scaling all individual exposures comprising a
tile to a common median scale; this step ensured that opposed edges of
a combined frame could be effectively used to photometrically tie down
adjacent tiles. This is a new feature of the pipeline.

All science images were mosaicked, aligned and median combined using
SWarp \citep{SWarp} to a plate scale of ${0.''}266$/pixel. Source
detection and photometry measurements for the science catalogs were
performed using SExtractor \citep{SEx}.
Large variations in the Galactic dust absorption on the 5hr field led
us to implement dust correction for every source in each observed band
in the science catalogs utilizing the infrared maps and C-routines
provided by \cite{Schlegel98}. This is particularly important to
obtain unbiased colors for photometric redshifts. In
Figure~\ref{fig:dustmap} we show our custom generated $E(B-V)$
Galactic dust absorption maps for the 5hr and 23hr regions, where
darker regions represent areas of higher obscuration. From this figure
we can see that in the 23hr region foreground Galactic
absorption is negligible, while at 5hr it can be as high as
$E(B-V)\simeq0.14$ magnitudes. Finally, we determine photometric
redshifts from the four-band optical images and the Galactic--extinction--corrected
magnitudes using BPZ \citep{BPZ} following the same procedure as
discussed in \cite{SCSI}.

\begin{deluxetable*}{cccccccccc}
%\tabletypesize{\footnotesize }  % for emulate apj
\tablecaption{2005-2007 Observations in the 23hr and 5hr Fields}
\tablehead{
\multicolumn{2}{c}{} & \multicolumn{4}{c}{\# of Tiles Obs 05h} 
                     & \multicolumn{4}{c}{\# of Tiles Obs 23h}\\
\colhead{Date} & 
\colhead{Photometric} & 
\colhead{$g$} & 
\colhead{$r$} &
\colhead{$i$} &
\colhead{$z$} &
\colhead{$g$} & 
\colhead{$r$} &
\colhead{$i$} &
\colhead{$z$} 
}
\startdata
18 Nov 2005 &   yes &  \nodata &  \nodata &  1.7 &  1.3  &  1.0 &  1.0 &  3.0 &  3.0 \\
19 Nov 2005 &   yes &  \nodata &  \nodata &  4.7 &  5.0  &  4.5 &  4.5 &  1.0 &  1.0 \\
20 Nov 2005 &    no &  \nodata &  \nodata &  5.7 &  5.7  &  8.5 &  8.0 &  \nodata &  \nodata \\
22 Nov 2005 &    no &  \nodata &  \nodata &  \nodata &  \nodata  &  \nodata &  \nodata &  \nodata &  \nodata \\
24 Nov 2005 &   yes &  9.0 &  9.0 &  1.7 &  1.7  &  5.5 &  5.5 &  \nodata &  \nodata \\
26 Nov 2005 &   yes &  9.5 &  9.0 &  1.0 &  1.0  &  4.5 &  4.5 &  1.0 &  1.0 \\
28 Nov 2005 &   yes &  4.0 &  3.0 &  4.0 &  4.0  &  \nodata &  \nodata &  2.7 &  2.7 \\
30 Nov 2005 &   yes & 12.5 & 12.5 &  \nodata &  \nodata  &  5.0 &  5.0 &  0.3 &  0.3 \\
02 Dec 2005 &   yes &  \nodata &  \nodata &  \nodata &  \nodata  &  3.5 &  3.5 &  \nodata &  \nodata \\
04 Dec 2005 &    no &  9.0 &  9.0 &  \nodata &  \nodata  &  2.0 &  2.0 &  \nodata &  \nodata \\
05 Dec 2005 &    no &  2.5 &  2.0 &  2.7 &  2.7  &  2.0 &  2.0 &  1.3 &  1.7 \\
06 Dec 2005 &   yes &  \nodata &  \nodata &  4.0 &  4.0  &  \nodata &  \nodata &  1.3 &  1.0 \\
08 Dec 2005 &   yes &  \nodata &  \nodata &  4.0 &  4.0  &  \nodata &  \nodata &  2.3 &  2.7 \\
10 Dec 2005 &   yes &  \nodata &  \nodata &  5.7 &  5.7  &  \nodata &  \nodata &  3.0 &  3.0 \\
11 Dec 2005 &   yes &  \nodata &  \nodata &  4.3 &  4.3  &  \nodata &  \nodata &  3.0 &  3.3 \\
23 Oct 2006 &   yes &  9.0 & 10.5 &  \nodata &  \nodata  &  \nodata &  \nodata &  5.0 &  5.0 \\
25 Oct 2006 &    no &  \nodata &  \nodata &  \nodata &  \nodata  &  \nodata &  \nodata &  \nodata &  \nodata \\
27 Oct 2006 &   yes & 10.5 &  7.0 &  \nodata &  \nodata  &  \nodata &  \nodata &  5.0 &  5.3 \\
28 Oct 2006 &    no &  \nodata &  \nodata &  \nodata &  \nodata  &  \nodata &  \nodata &  5.3 &  5.0 \\
29 Oct 2006 &   yes &  8.0 &  8.0 &  2.0 &  2.0  &  \nodata &  \nodata &  4.0 &  4.3 \\
30 Oct 2006 &   yes &  \nodata &  \nodata &  \nodata &  \nodata  &  \nodata &  \nodata &  5.0 &  5.3 \\
31 Oct 2006 &   yes &  \nodata &  \nodata &  6.0 &  6.0  &  \nodata &  \nodata &  3.0 &  2.7 \\
12 Dec 2006 &   yes &  \nodata &  \nodata &  3.0 &  1.7  &  \nodata &  \nodata &  \nodata &  \nodata \\
13 Dec 2006 &   yes &  \nodata &  \nodata &  3.0 &  3.3  &  1.0 &  0.5 &  \nodata &  \nodata \\
14 Dec 2006 &    no &  \nodata &  \nodata &  2.7 &  2.7  &  \nodata &  \nodata &  \nodata &  \nodata \\
15 Dec 2006 &   yes &  4.0 &  4.0 &  5.7 &  5.7  &  \nodata &  \nodata &  \nodata &  \nodata \\
16 Dec 2006 &    no &  \nodata &  \nodata &  4.0 &  3.7  &  \nodata &  \nodata &  \nodata &  \nodata \\
17 Dec 2006 &   yes &  7.0 &  6.5 &  4.3 &  5.0  &  \nodata &  \nodata &  \nodata &  \nodata \\
18 Dec 2006 &   yes &  8.0 &  8.5 &  \nodata &  \nodata  &  \nodata &  \nodata &  \nodata &  \nodata \\
19 Dec 2006 &   yes &  3.0 &  3.0 &  6.3 &  6.3  &  \nodata &  \nodata &  \nodata &  \nodata \\
20 Dec 2006 &   yes &  \nodata &  \nodata &  4.0 &  4.0  &  \nodata &  \nodata &  \nodata &  \nodata \\
21 Dec 2006 &   yes &  \nodata &  \nodata &  8.7 &  8.7  &  \nodata &  \nodata &  \nodata &  \nodata \\
22 Dec 2006 &   yes &  3.5 &  3.5 &  2.3 &  2.0  &  \nodata &  \nodata &  \nodata &  \nodata \\
23 Dec 2006 &   yes &  6.5 &  6.0 &  6.0 &  6.3  &  \nodata &  \nodata &  \nodata &  \nodata \\
11 Sep 2007 &   yes &  2.5 &  3.0 &  2.3 &  2.7  &  4.5 &  5.0 &  3.3 &  4.0 \\
12 Sep 2007 &   yes &  \nodata &  \nodata &  \nodata &  \nodata  &  1.0 &  1.0 &  1.0 &  1.0 \\
13 Sep 2007 &   yes &  \nodata &  \nodata &  \nodata &  \nodata  &  1.5 &  1.0 &  1.3 &  1.3 \\
14 Sep 2007 &   yes &  1.5 &  1.5 &  1.7 &  1.3  &  4.5 &  5.0 &  4.7 &  4.7 \\
15 Sep 2007 &    no &  2.0 &  2.0 &  0.7 &  1.3  &  2.5 &  2.5 &  1.0 &  1.0 \\
16 Sep 2007 &   yes &  \nodata &  \nodata &  \nodata &  \nodata  &  2.5 &  2.5 &  \nodata &  \nodata \\
17 Sep 2007 &   yes &  1.0 &  \nodata &  \nodata &  \nodata  &  \nodata &  \nodata &  2.0 &  1.7 \\
18 Sep 2007 &   yes &  \nodata &  \nodata &  2.3 &  2.0  &  3.0 &  2.5 &  7.0 &  6.7 \\
19 Sep 2007 &   yes &  \nodata &  \nodata &  \nodata &  \nodata  &  \nodata &  \nodata &  4.7 &  5.0 \\
11 Oct 2007 &   yes &  \nodata &  \nodata &  \nodata &  \nodata  &  \nodata &  \nodata &  \nodata &  \nodata \\
12 Oct 2007 &    no &  \nodata &  \nodata &  \nodata &  \nodata  &  \nodata &  \nodata &  \nodata &  \nodata \\
30 Oct 2007 &    no &  \nodata &  \nodata &  1.0 &  \nodata  &  \nodata &  0.5 &  \nodata &  \nodata \\
31 Oct 2007 &    no &  \nodata &  \nodata &  \nodata &  \nodata  &  \nodata &  \nodata &  \nodata &  \nodata \\
01 Nov 2007 &   yes &  \nodata &  \nodata &  1.3 &  2.3  &  2.5 &  2.0 &  \nodata &  \nodata \\
02 Nov 2007 &   yes &  \nodata &  \nodata &  \nodata &  \nodata  &  5.0 &  5.0 &  \nodata &  \nodata \\
03 Nov 2007 &   yes &  \nodata &  \nodata &  3.0 &  3.0  &  9.5 &  9.5 &  \nodata &  \nodata \\
04 Nov 2007 &   yes &  5.0 &  5.5 &  \nodata &  \nodata  &  3.0 &  3.0 &  \nodata &  \nodata \\
06 Nov 2007 &   yes &  \nodata &  \nodata &  \nodata &  \nodata  &  \nodata &  \nodata &  4.7 &  4.3 \\
07 Nov 2007 &   yes &  1.0 &  2.0 &  0.7 &  0.7  &  1.0 &  1.0 &  1.7 &  2.0 \\
08 Nov 2007 &   yes &  \nodata &  \nodata &  \nodata &  \nodata  &  \nodata &  \nodata &  \nodata &  \nodata \\
%% 119.0 115.5 110.333333333 110.333333333	%% 78.0 77.0 77.6666666667 77.6666666667
\enddata
\label{tab:obs}
\tablecomments{Observing conditions during the 2005-2007 runs of the
  BCS as well as the number of tiles observed on each night. We only
  show the dates in which tiles in the 23hr or 5hr contiguous regions were
  observed.}
\end{deluxetable*}

\subsection{Finding Clusters and Defining Membership}

In this section we describe our efforts to select clusters of galaxies
from multi-wavelength optical imaging. To this objective we follow an
identical procedure as the one we laid out in \cite{SCSI} as part of
our initial study of $8$~deg$^2$ using the 2005 observations at
23hr. We refer the reader there for an expanded and formal description
of the method as well as for simulations that address the
contamination and completeness of the search method. Here, we expanded
our procedure to the full 70~deg$^2$ of Blanco $griz$ contiguous
imaging.
Our cluster search algorithm uses a matched filter approach based 
on the one described in \cite{Postman96} to identify significant
overdensities. We define membership and estimate richness for
each peak in the overdensity maps using the MaxBCG prescription \citep{MaxBCG}.
Our method folds in the contributions from: a) a cluster spatial profile
filter function, b) a luminosity weight and c) the BPZ redshift probability
distribution from each source to generate likelihood density maps or a
``filtered'' galaxy catalog over the area covered by the survey as a
function of redshift. We generate likelihood density maps with a
constant pixel scale of $1.2$~arcmin at $\Delta z=0.1$ intervals
between $0.1<z<0.8$ over the surveyed regions. In
Figure~\ref{fig:densemap} we show an example of a likelihood density
map centered at $z=0.2$ on which we superpose outlines of the 78 and
112 tiles that define the 23hr and 5hr regions studied.

% New added for the referee
Because the mass of a cluster is not a direct observable, some
observable proxy for mass needs to be used in order to obtain
estimates for a given cluster sample. Such proxies include the X-ray
flux and temperature \citep{HIFLUGCS,Krastov06,
  Rykoff08b,Rozo-X,Vikhlinin09}, weak lensing shear
\citep{Sheldon09,Okabe10} and optical galaxy richness of clusters
\citep{Becker07,Rozo-N}.
For this analysis we use the latest mass tracers for clusters of
galaxies that are based on optically observed parameters
\citep{Johnston07,Reyes08} extracted from a sample of around $13,000$
clusters from the Sloan Digital Sky Survey (SDSS) MaxBCG catalog
\citep{MaxBCG}. To use these mass tracers we define membership in
similar fashion as \citet{MaxBCG}. We then visually inspected each
candidate-cluster peak in the density maps and selected the brightest
elliptical galaxy in the cluster (BCG), which was taken to be the
center and initial redshift of the system.  We then use galaxies
photometrically classified as E or E/S0s according to their BPZ
spectral type and within a projected radius of $0.5h^{-1}$ Mpc and
redshift interval $|z-z_o| = |\Delta z|=0.05$ to obtain a local
color-magnitude relation for each color combination as well as the
cluster mean redshift, $z_c$, for all cluster members, using a
$3\sigma$ median sigma-clipping algorithm. We use these to determine
$N_{\rm 1Mpc}=N_{\rm gal}$, the number of galaxies within $1h^{-1}$Mpc
of the cluster center. For our richness measurements we estimated the
galaxy background contamination and implemented an appropriate
background subtraction method following the same procedure described
in \cite{SCSI} (see section 3.1).
We use a statistical removal of unrelated field galaxies with similar
colors and redshifts that were projected along the line of sight to
each cluster. We estimate the surface number density of ellipticals in
an annulus surrounding the cluster (within $R_{200}<r<2 R_{200}$) with
$\Delta z=0.05$ and the same colors as the cluster members. We measure
this background contribution around the outskirts of each cluster and
obtain a corrected value $N_{gal}$ which is used to compute $R_{200}$
and then corresponding values of \n200 and $L_{200}$. The magnitude of
the correction ranges between $15-20$\%. We will refer to the
corrected values hereafter.

\begin{figure*}
%% \begin{figure} % -- MS style
%% \centerline{
%% \includegraphics[width=2.6in]{f5a.eps}
%% \includegraphics[width=2.6in]{f5b.eps}}
%% \centerline{
%% \includegraphics[width=2.6in]{f5c.eps}
%% \includegraphics[width=2.6in]{f5d.eps}}
%% \centerline{
%% \includegraphics[width=2.6in]{f5e.eps}
%% \includegraphics[width=2.6in]{f5f.eps}}
\centerline{
\includegraphics[width=3.1in]{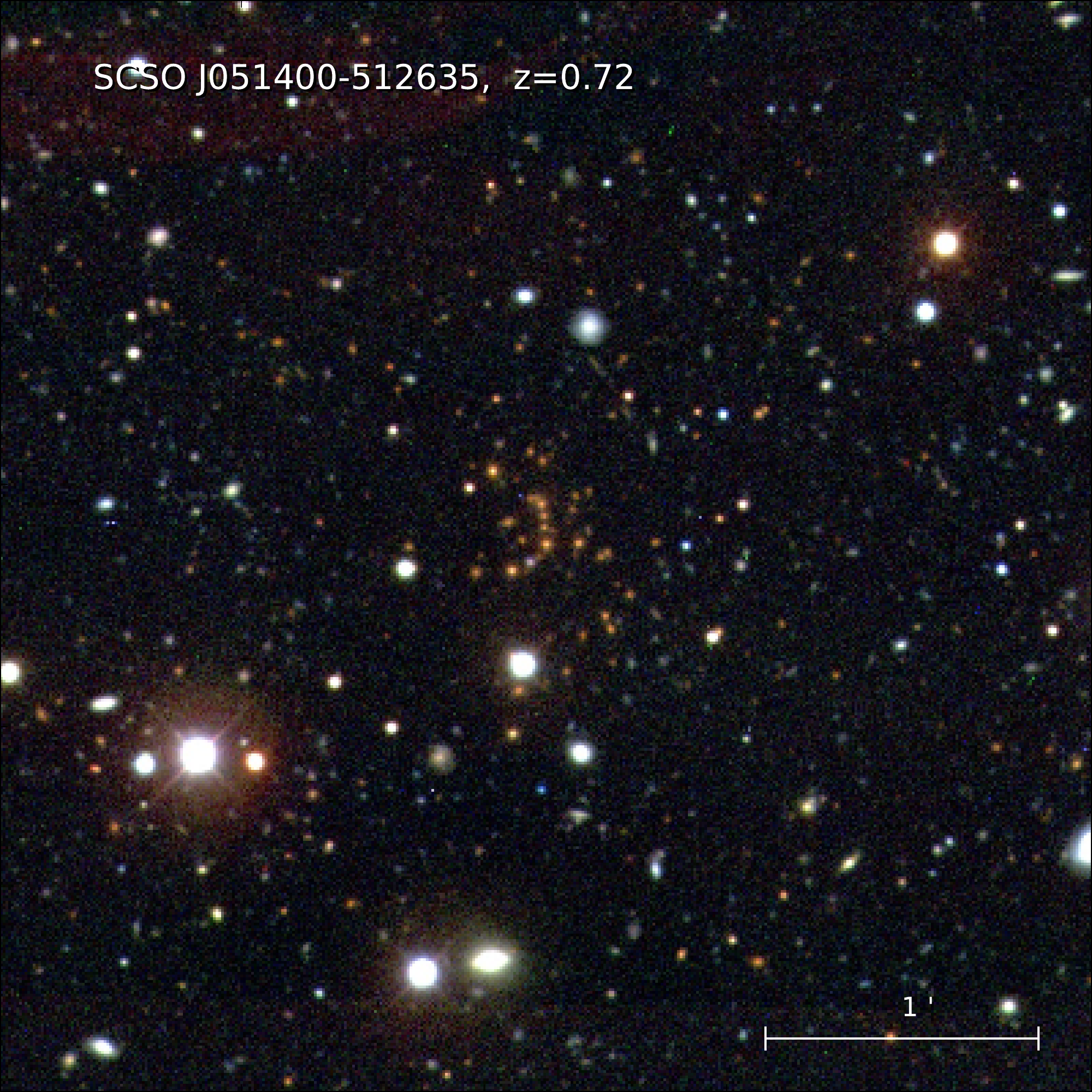}
\includegraphics[width=3.1in]{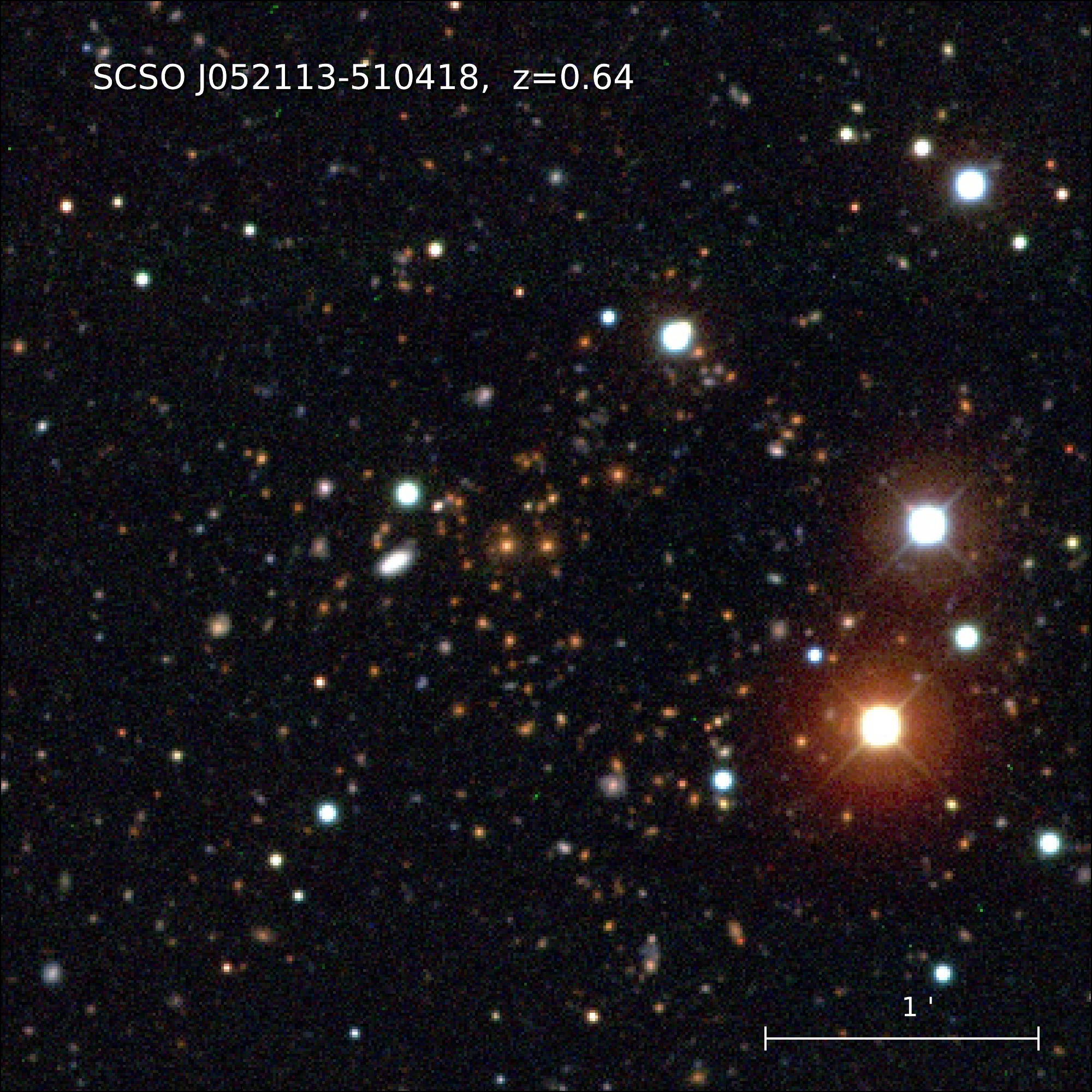}}
\centerline{
\includegraphics[width=3.1in]{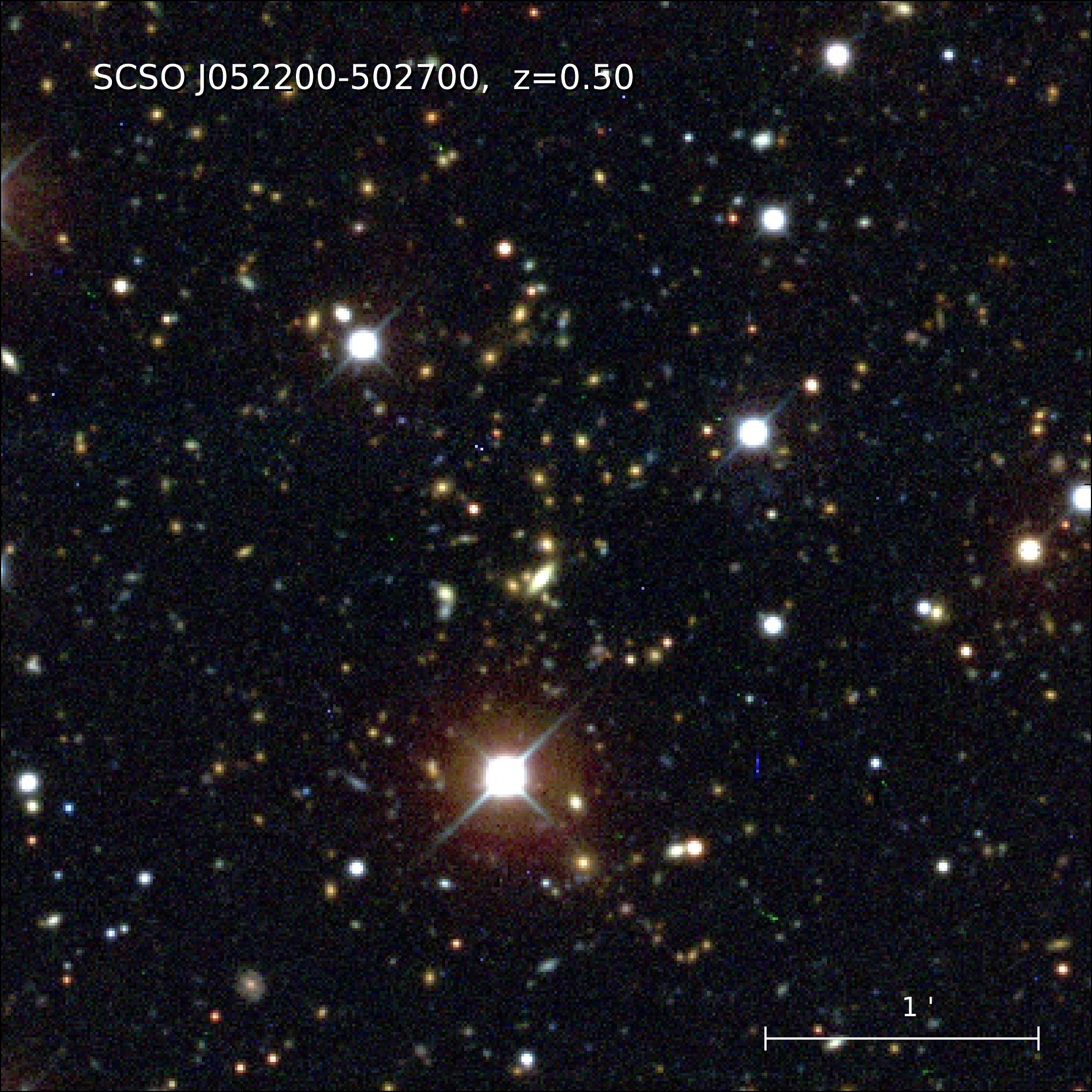}
\includegraphics[width=3.1in]{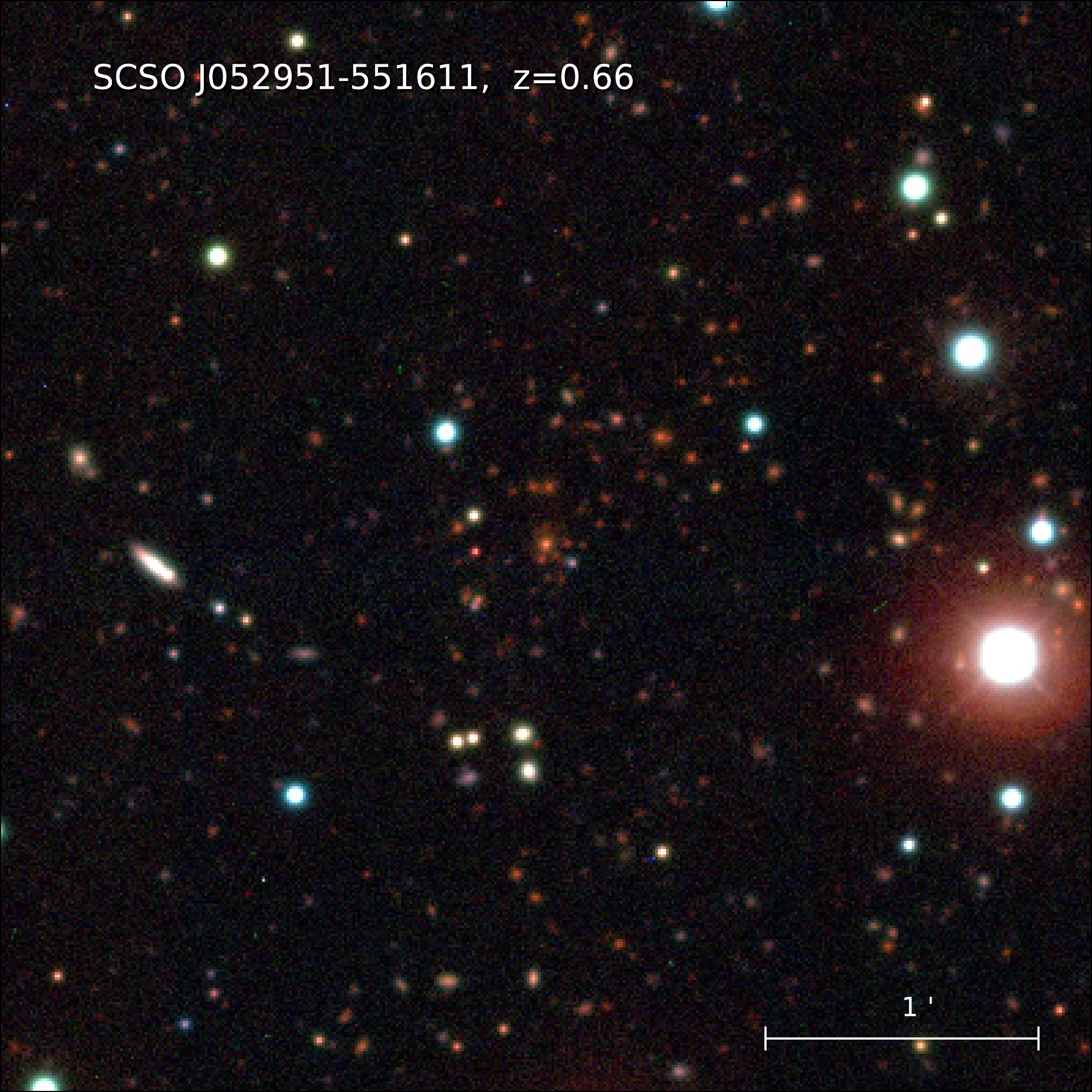}}
\centerline{
\includegraphics[width=3.1in]{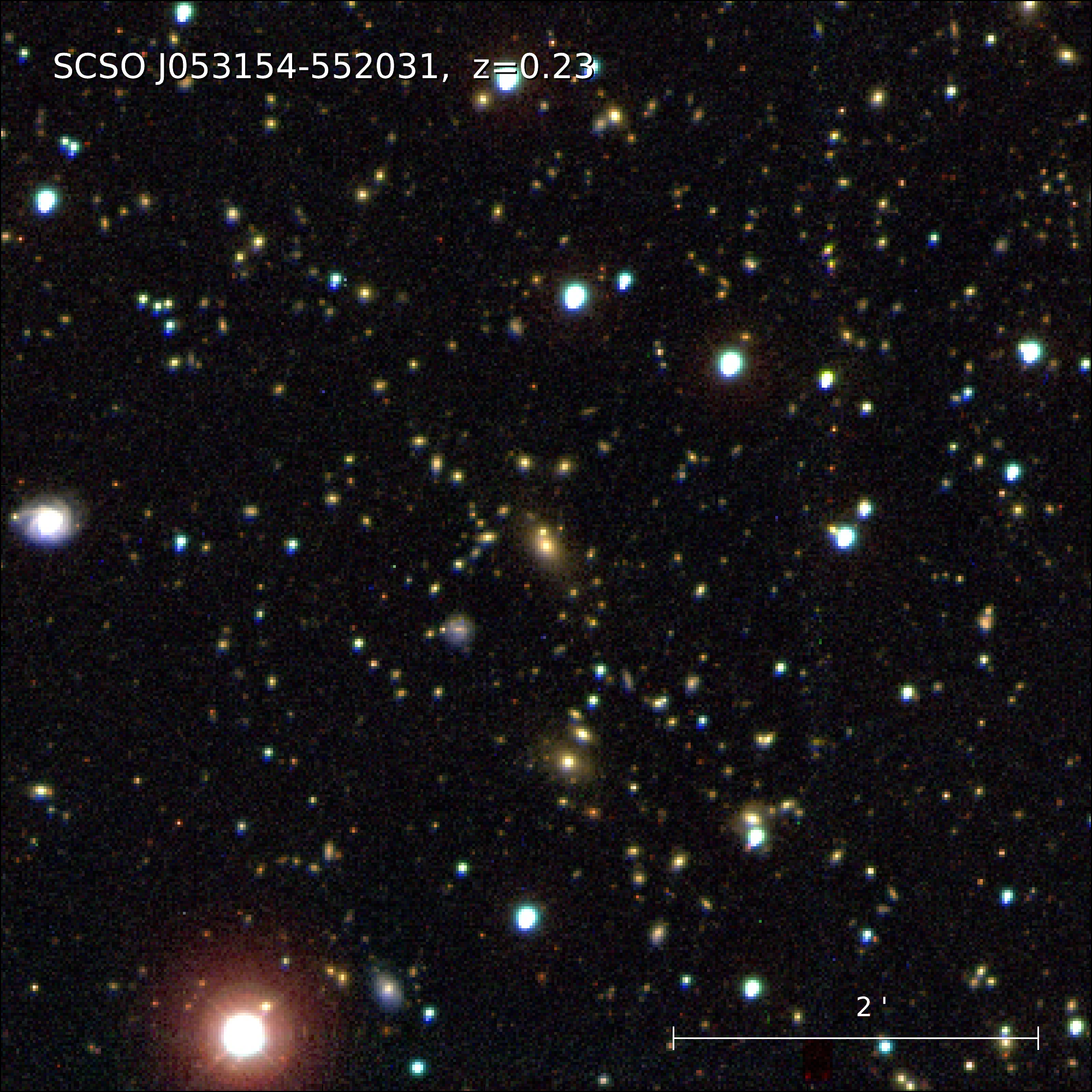}
\includegraphics[width=3.1in]{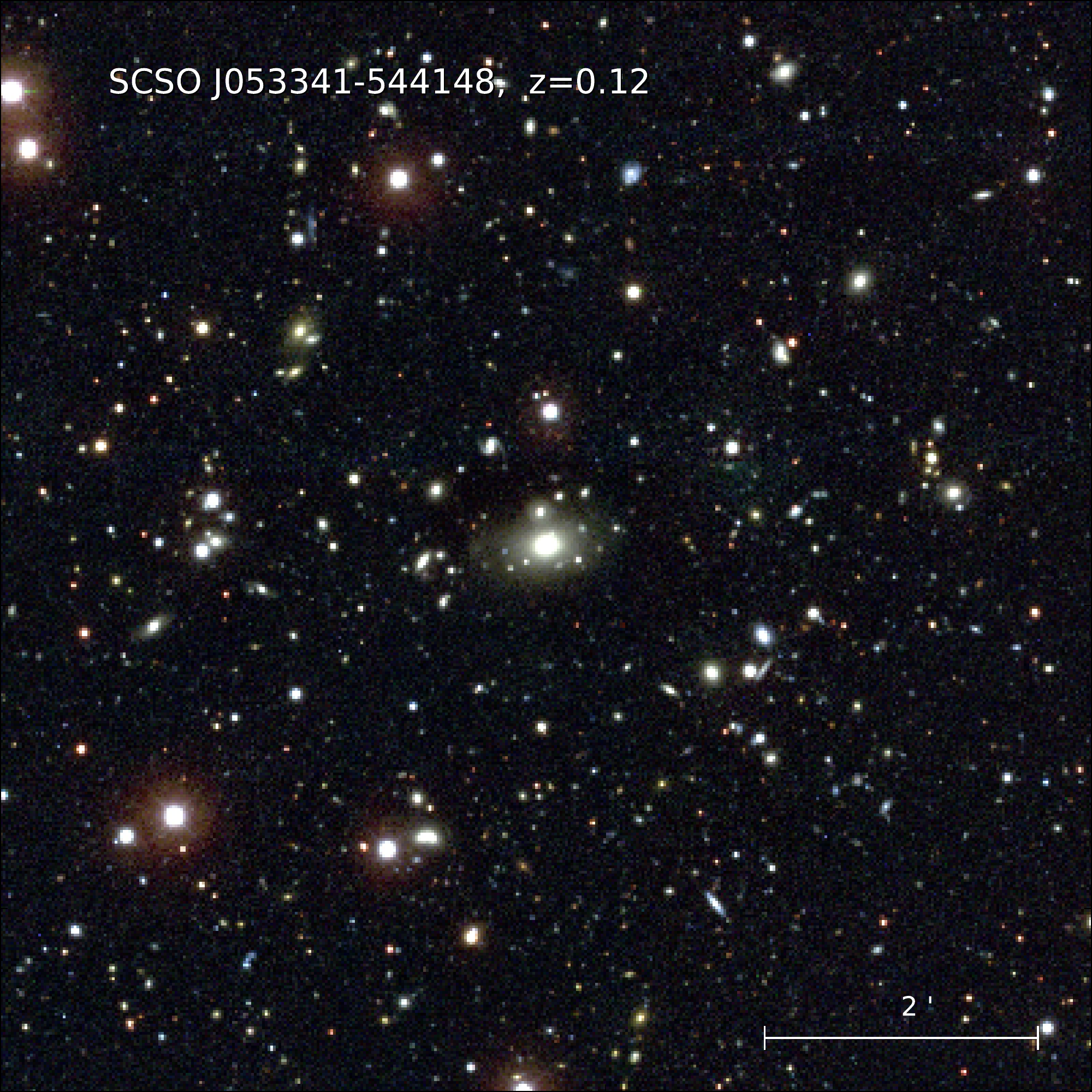}}
\caption{Composite $gri$ color image for nine newly discovered massive
  SCS clusters in the 5hr field. We indicate the cluster's redshift
  using the median value from galaxy members within $250$~kpc of the
  cluster's center.}
\label{fig:clusters05hr}
\end{figure*}

\begin{figure}
\includegraphics[width=3.0in]{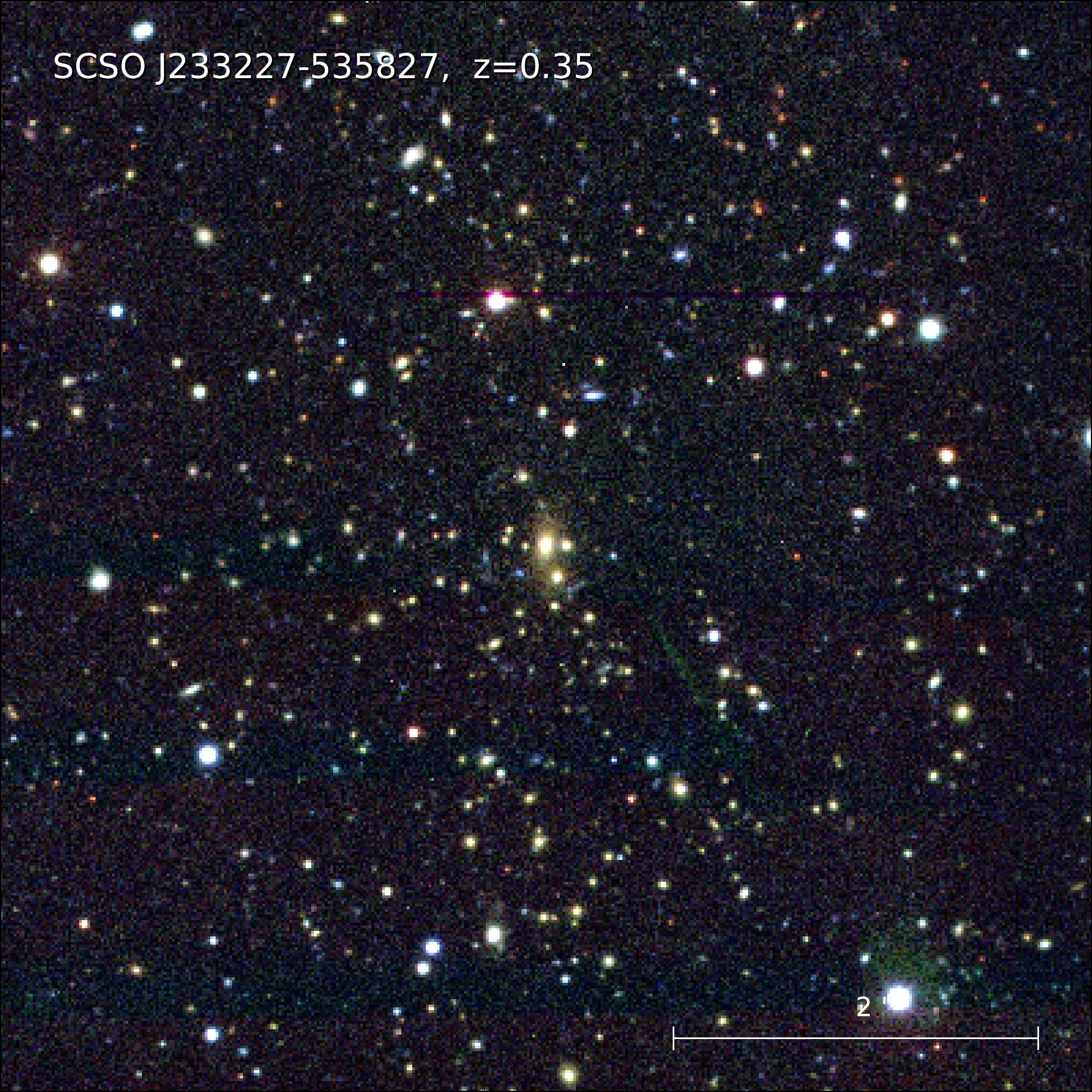}\\
\includegraphics[width=3.0in]{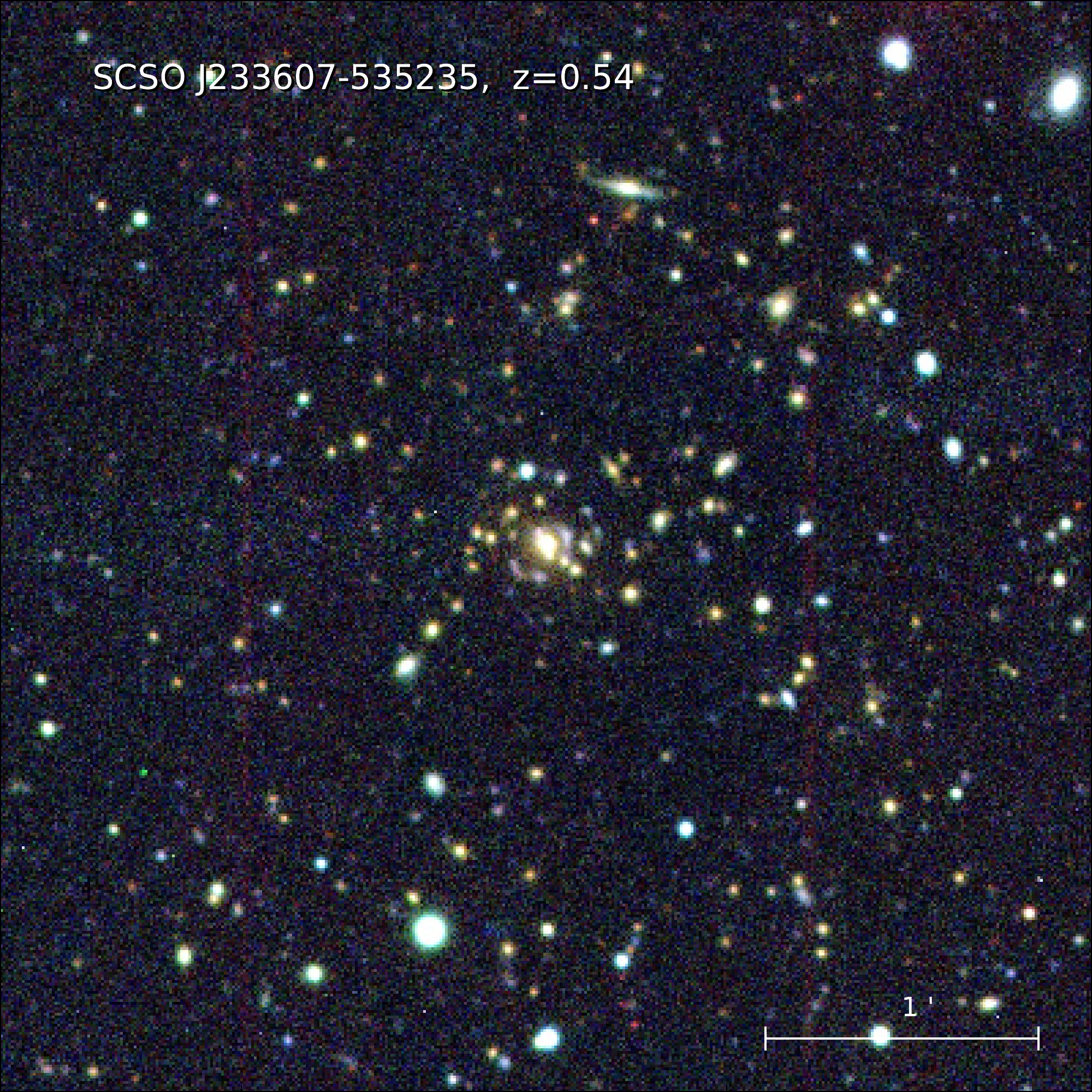}\\
\includegraphics[width=3.0in]{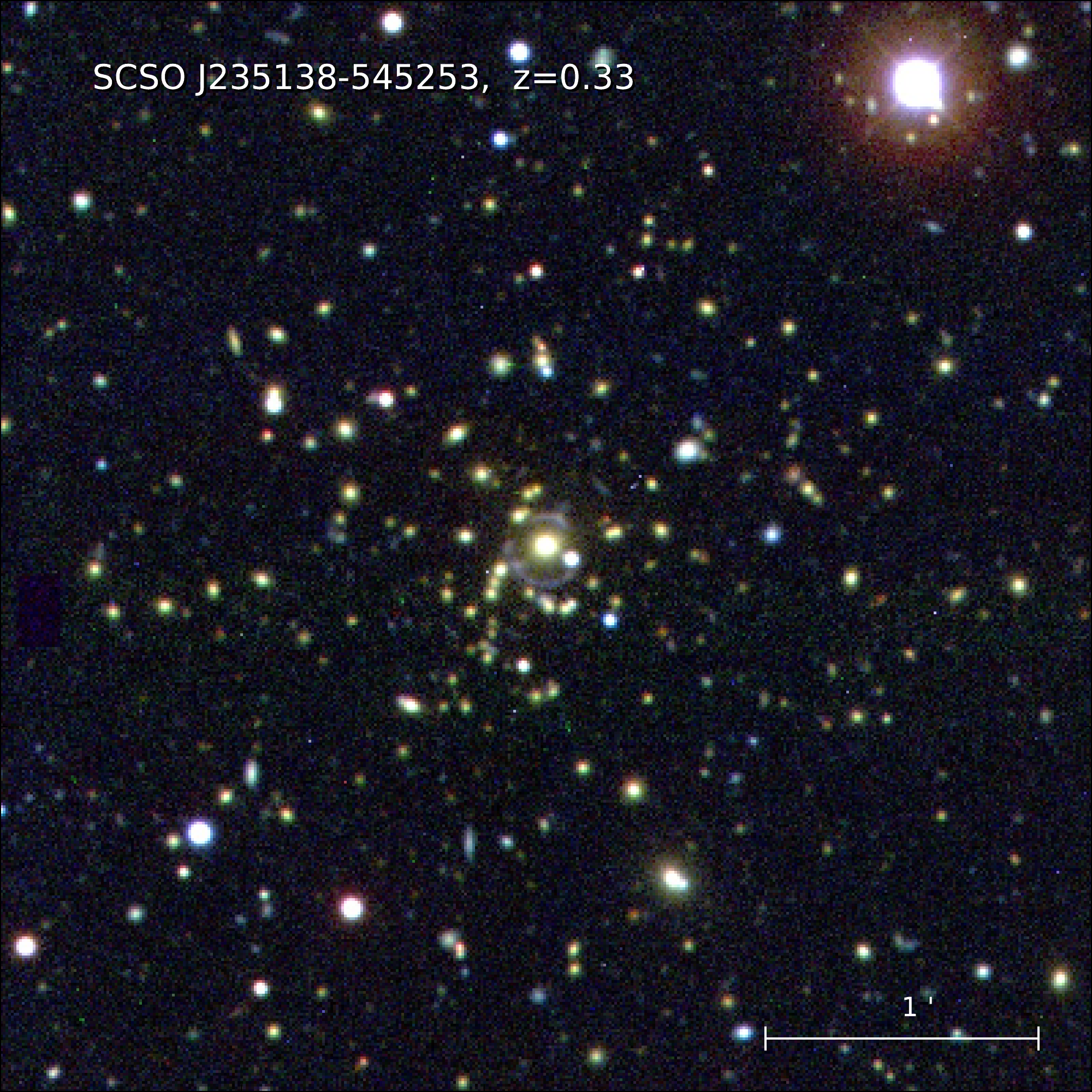}\\
\caption{Composite $gri$ color image for nine newly discovered massive
  SCS clusters in the 23hr field. We indicate the cluster's redshift
  using the median value from galaxy members within $250$~kpc of the
  cluster's center.}
\label{fig:clusters23hr}
\end{figure}

% Nice color figures
Finally we show, as examples of the depth and data quality, composite
$gri$ color images of several clusters in the 5hr and 23hr regions in
Figures~\ref{fig:clusters05hr} and \ref{fig:clusters23hr},
respectively, that cover a wide range of redshifts. 

\section{Results}

\subsection{Optical Cluster Masses}

The observational quantities required as input to the cluster mass
scaling relation are \n200, \L200, and \LBCG. The cluster richness,
\n200, is the number of E and E/S0 galaxies within a given radius, originally defined
by \cite{Hansen-05} as $R_{200}= 0.156N_{\rm  1Mpc}^{0.6}h^{-1}$~Mpc,
with colors and luminosities that satisfy specific conditions for
membership. Similarly, $L_{200}$ is the total rest-frame
integrated $r$-band luminosity, $k$-corrected to $z=0.25$, of all
member galaxies included in \n200, and $L_{\rm BCG}$ is the similarly
defined rest-frame $r$-band luminosity of the BCG. \cite{Reyes08}
provide power-law functions for both the luminosity-mass and
richness-mass relations (see section 5.2.1 in their paper).

Both \cite{Johnston07} and \cite{Reyes08} found that the
luminosity-mass and richness-mass relations were well described by
power-law functions and they measured the normalizations and slopes in
these relations using $\chi^2$ minimization. 
We computed the two fitting functions based on $L_{200}$ and \n200,
(see section 5.2.1 from \citealt{Reyes08} for full details), which are
described as:
\begin{equation}
M(N_{200},L_{\rm BCG}) = M_N^0(N_{200}/20)^{\alpha_N}(L_{\rm BCG}/\bar{L}_{\rm BCG}^{(N)})^{\gamma_N} 
\label{eq:M1}
\end{equation}
and
\begin{equation}
M(L_{200},L_{\rm BCG}) = M_L^0(L_{200}/40)^{\alpha_L}(L_{\rm BCG}/\bar{L}_{\rm BCG}^{(L)})^{\gamma_L}
\label{eq:M2}
\end{equation}
where $M$ is the mass observational equivalent of $M_{200\bar{\rho}}$
(i.e.~the halo mass enclosed within a radius of spherical volume
within which the mean density is 200 times the average density) in
units of $10^{14}M_{\sun}$, $L_{200}$ is in units of
$10^{10}h^{-2}L_{\sun}$ and the \LBCG\ dependence is normalized by its
mean value. This is also described by a power-law function for a given
value of \L200\ and \n200:
\begin{equation}
\bar{L}_{\rm BCG}^{(N)} \equiv \bar{L}_{\rm BCG}(N_{200}) = a_N N_{200}^{b_N}
\label{eq:L1}
\end{equation}
and
\begin{equation}
\bar{L}_{\rm BCG}^{(L)} \equiv \bar{L}_{\rm BCG}(L_{200}) = a_L L_{200}^{b_L}
\label{eq:L2}
\end{equation}

% About Reyes parameters
The published best-fitting parameters for $M^0$, $\alpha$ and $\gamma$
in Eqs.~(\ref{eq:M1}) and (\ref{eq:M2}) as well as the new
erratum-corrected values of $a$ and $b$ (R.~Reyes, private
communication) for Eqs.~(\ref{eq:L1}) and (\ref{eq:L2}) are shown in
Table~\ref{tab:pars}. These recent changes in the values of the $a$
and $b$ parameters in \citet{Reyes08} have implications for our mass
estimation.  Specifically, the changes translate into a decrease in
mass when compared to our previous analysis of \cite{SCSI} which used
the initial parameters from their pre-print paper. As we discuss
below, this change also affects the recovery of clusters in
\cite{SCSI} for the sky region that overlaps with this study.

\begin{figure}
%\centerline{\includegraphics[width=6.0in]{f3.pdf}}
\centerline{\includegraphics[width=4.0in]{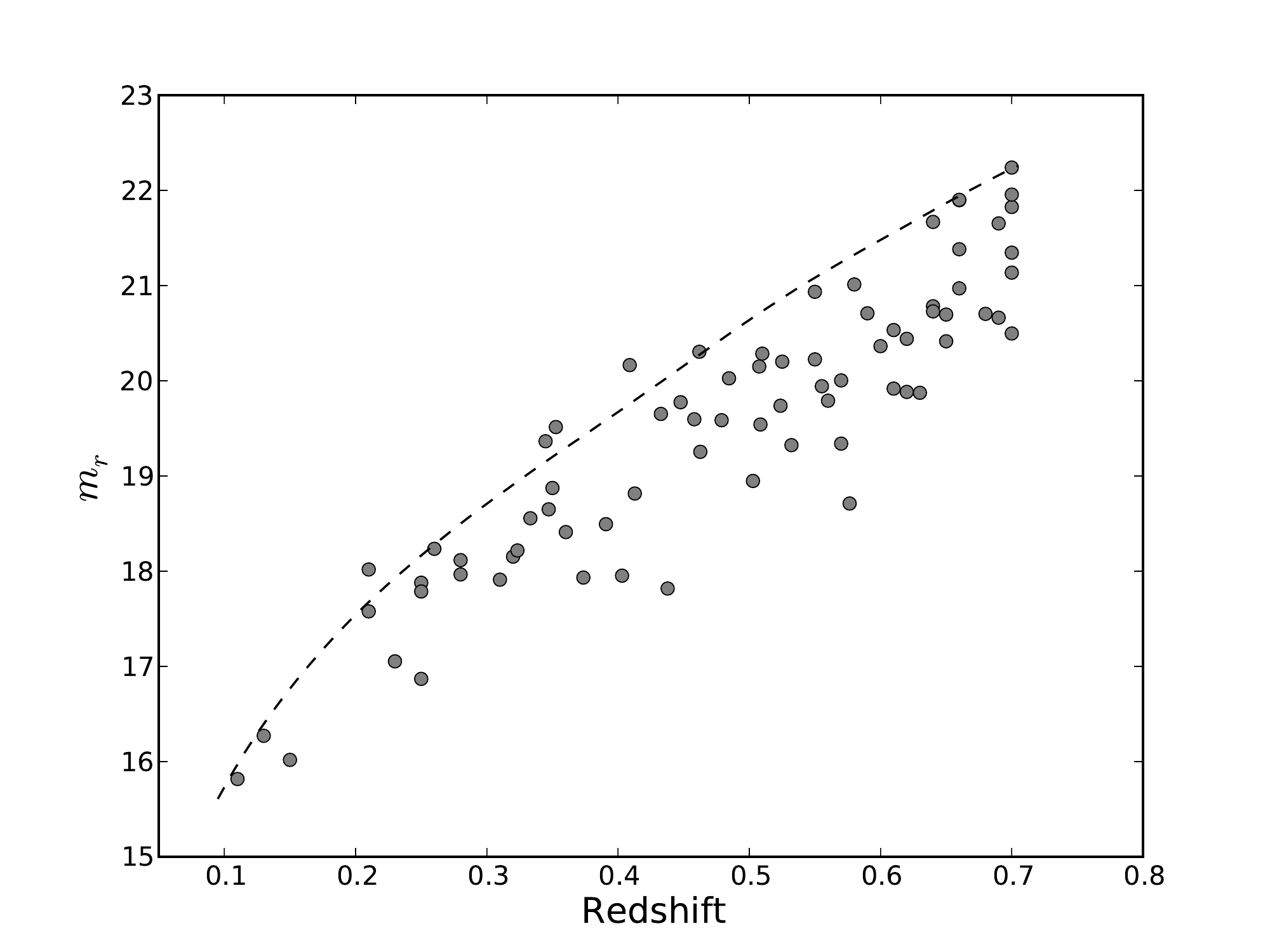}}
\caption{The $r-$band observed magnitude for the BCGs in our sample as
  a function of the redshift of the cluster (filled circles). We also
  show the $M^*  - 1.5$ BCG-redshift relationship from \cite{LS06}
  as the dashed curve.}
\label{fig:magz}
\end{figure}

% About The tables
We use this parametrization to obtain the optical mass estimates
\MN200 and \ML200, based on \n200 and \L200, respectively, for all of
the visually inspected clusters. Based on these estimates we defined
our catalog of massive clusters by selecting all systems with either
$M(N_{200})>3\times 10^{14}M_{\sun}$ or $M(L_{200})>3\times
10^{14}M_{\sun}$ and $N_{\rm gal} > 10$. This defines the sample: 61
systems in the 5hr region and 44 in the 23hr region for a total of 105
clusters over $70$~deg$^2$ of multi-band imaging. 
% Added for the referee
This mass threshold is aimed to include the upcoming $z<0.8$
significant SZE detections from SPT and ACT.
In Tables~\ref{tab:clustes23hr} and \ref{tab:clustes05hr} we display
the optical properties, photometric redshifts, positions and mass
estimates for all clusters at 23hr and 5hr respectively. In these
tables we provide for each cluster the photometric redshift of the BCG
as well as the mean photometric redshift for the system which was
estimated using galaxies within 250~kpc of the BCG.

% BCG Luminosities
As we discussed in \cite{MenanteauHughes09} the expected luminosity
range of BCGs in rich clusters has been already observationally
established from the SDSS \citep{LS06,MaxBCG} and it can be used as an
additional probe to confirm the presence of a massive cluster. To this
end in Figure~\ref{fig:magz} we compare the observed $r-$band
magnitudes of all BCGs in our sample as a function of redshift to a
parametrization of the observed $r-$band of SDSS BCGs (dashed
curve). This corresponds to the $M^*-1.5$ prescription from \cite{LS06}
where we have used $M^*$ from \cite{Blanton03} and allowed it to evolve
passively with redshift. We see in the figure that all of the
sources lie very close or well below (i.e., are intrinsically
brighter) than the model curve as we would expect for the BCG in a rich
cluster.

% Comparison with SCSI clusters.
In our previous study \citep{SCSI} we presented a similar cluster
analysis and mass estimation over $8$~deg$^2$ area in the 23hr region
which is fully contained in the current 70~deg$^2$ dataset. However,
due to the changes in the mass parametrization prescription we did not
recover all the massive clusters from \cite{SCSI} even though the
nominal mass threshold was the same ($M>3\times10^{14}M_{\sun}$) as
here. Therefore, only 3 out of 8 clusters from Table~ 5 of \cite{SCSI}
are massive enough to appear in the current paper's
Table~\ref{tab:clustes23hr}, while the remaining clusters, although
detected, now have masses that fall below the
$3\times10^{14}~M_{\sun}$ mass limit. Similarly, when comparing with
the weak lensing mass estimates of \cite{McInnes09} we can only match
the same 3 sources with clusters in their analysis (see Table~2 of
their paper). On the other hand, there is now better agreement between the updated
$M(L_{200})$ values and the weak lensing masses for the matched
clusters.

\subsection{Lensing Rate}
Out of the 105
massive clusters in the sample we report two systems (see the two
lower panels of Figure~\ref{fig:clusters23hr}) with obvious evidence
of arcs near the BCG. Of these two lensing clusters,
SCSO~235138$-$545253 at $z=0.33$ is a potentially unique system in
that it shows a large diameter ($\sim15''$), nearly complete Einstein
Ring embedded in a massive cluster.  Although our density of lensing
clusters is not widely different from the \cite{Gladders03} RCS sample
of 5 lensing clusters over $\sim90$~deg$^2$, all of the Gladders
et~al. clusters are at $z>0.64$, while both of our systems are at
$z<0.54$ making them completely exclusive in redshift. Moreover, none
of the \cite{Gladders03} sample shows anything close to a complete
Einstein Ring.

\begin{deluxetable*}{rrrrrrrrrrr}[b]
%\begin{deluxetable}{rrrrrrrrrrr} %% -- for MS-style
%\tabletypesize{\footnotesize } 
%\rotate
\tablecaption{Mass-richness power-law function Best fitting parameters
from \cite{Reyes08}}
\tablehead{
\multicolumn{1}{r}{} & \multicolumn{2}{r}{$(10^{10}h^{-2}L_{\sun})$}\\
\colhead{Redshift} & 
\colhead{$a_N$} & 
\colhead{$a_L$} & 
\colhead{$b_N$} & 
\colhead{$b_L$} &
\colhead{$M_N^0$} &
\colhead{$\alpha_N$} &
\colhead{$\gamma_N$} &
\colhead{$M_L^0$} &
\colhead{$\alpha_L$} &
\colhead{$\gamma_L$} 
}
\startdata
$0.10<z<0.23$ & 1.54   & 0.61  & 0.41 & 0.67 & $1.27\pm0.08$ & $1.20\pm0.09$ & $0.71\pm 0.14$ & $1.81\pm0.15$ & $1.27\pm0.17$ & $0.40\pm0.23$ \\ 
$0.23<z<0.70$ & 1.64   & 0.58  & 0.43 & 0.66 & $1.57\pm0.14$ & $1.12\pm0.15$ & $0.34\pm 0.24$ & $1.76\pm0.22$ & $1.30\pm0.29$ & $0.26\pm0.41$ \\ 
\enddata
\label{tab:pars}
\end{deluxetable*}
%\end{deluxetable} % -- MS style

%%%%%%%%%%%%%%%%%%%%%%%%%%
%% 23hr clusters Table %%%
%%%%%%%%%%%%%%%%%%%%%%%%%%
\begin{deluxetable*}{rrrccrrrr}
%\begin{deluxetable}{rrrccrrrr} % -- MS style
%\tabletypesize{\footnotesize } % -- MS style
%\rotate
\tablecaption{Optical Clusters with $M(N_{200})>3\times
  10^{14}M_{\sun}$ or $M(L_{200})>3\times
  10^{14}M_{\sun}$ in the 23hr field}
\tablewidth{0pt}
\tablehead{
\colhead{ID} & 
\colhead{$z_{\rm cluster}$} & 
\colhead{$z_{\rm BCG}$} & 
\colhead{$N_{\rm gal}$} &
\colhead{\n200} &
\colhead{$L_{200}[L_{\sun}]$} &
\colhead{$M(N_{200})$} &
\colhead{$M(L_{200})$} \\
\multicolumn{6}{c}{} & \multicolumn{2}{c}{$(M_{\sun})$}  
}
\startdata
SCSO~J231340$-$521919 & 0.21 & 0.21 & $ 61.2\pm  7.8$ & $124.8\pm 11.2$ & $ 2.0 \times 10^{12}\pm2.3 \times 10^{11}$ & $3.1 \times 10^{14}$ & $2.9 \times 10^{14}$  \\
SCSO~J231403$-$564710 & 0.60 & 0.60 & $ 22.7\pm  4.8$ & $ 22.4\pm  4.7$ & $ 1.2 \times 10^{12}\pm5.2 \times 10^{10}$ & $2.2 \times 10^{14}$ & $3.5 \times 10^{14}$  \\
SCSO~J231455$-$555308 & 0.21 & 0.23 & $ 59.6\pm  7.7$ & $ 79.9\pm  8.9$ & $ 2.0 \times 10^{12}\pm2.2 \times 10^{11}$ & $4.6 \times 10^{14}$ & $4.7 \times 10^{14}$  \\
SCSO~J231511$-$523322 & 0.36 & 0.39 & $ 29.0\pm  5.4$ & $ 33.6\pm  5.8$ & $ 2.3 \times 10^{12}\pm1.0 \times 10^{11}$ & $3.9 \times 10^{14}$ & $8.2 \times 10^{14}$  \\
SCSO~J231629$-$554535 & 0.51 & 0.53 & $ 33.9\pm  5.8$ & $ 38.1\pm  6.2$ & $ 2.8 \times 10^{12}\pm1.1 \times 10^{11}$ & $5.2 \times 10^{14}$ & $1.1 \times 10^{15}$  \\
SCSO~J231651$-$545356 & 0.36 & 0.36 & $ 35.5\pm  6.0$ & $ 40.3\pm  6.3$ & $ 9.6 \times 10^{11}\pm9.6 \times 10^{10}$ & $3.9 \times 10^{14}$ & $2.7 \times 10^{14}$  \\
SCSO~J231717$-$565723 & 0.74 & 0.73 & $ 21.0\pm  4.6$ & $ 21.8\pm  4.7$ & $ 1.4 \times 10^{12}\pm5.2 \times 10^{10}$ & $1.7 \times 10^{14}$ & $3.3 \times 10^{14}$  \\
SCSO~J231848$-$561711 & 0.51 & 0.51 & $ 28.8\pm  5.4$ & $ 31.4\pm  5.6$ & $ 1.6 \times 10^{12}\pm9.4 \times 10^{10}$ & $3.4 \times 10^{14}$ & $5.2 \times 10^{14}$  \\
SCSO~J231930$-$563858 & 0.36 & 0.37 & $ 35.4\pm  6.0$ & $ 38.4\pm  6.2$ & $ 1.1 \times 10^{12}\pm8.2 \times 10^{10}$ & $3.5 \times 10^{14}$ & $2.9 \times 10^{14}$  \\
SCSO~J232001$-$565222 & 0.80 & 0.80 & $ 18.0\pm  4.2$ & $ 13.8\pm  3.7$ & $ 1.3 \times 10^{12}\pm5.4 \times 10^{10}$ & $1.8 \times 10^{14}$ & $4.6 \times 10^{14}$  \\
SCSO~J232423$-$565705 & 0.75 & 0.75 & $ 25.0\pm  5.0$ & $ 30.6\pm  5.5$ & $ 2.5 \times 10^{12}\pm8.0 \times 10^{10}$ & $2.3 \times 10^{14}$ & $6.5 \times 10^{14}$  \\
SCSO~J232437$-$530047 & 0.73 & 0.75 & $ 38.7\pm  6.2$ & $ 63.9\pm  8.0$ & $ 2.2 \times 10^{12}\pm2.1 \times 10^{11}$ & $3.9 \times 10^{14}$ & $4.7 \times 10^{14}$  \\
SCSO~J232529$-$532420 & 0.74 & 0.71 & $ 40.8\pm  6.4$ & $ 46.5\pm  6.8$ & $ 2.2 \times 10^{12}\pm2.3 \times 10^{11}$ & $2.6 \times 10^{14}$ & $4.5 \times 10^{14}$  \\
SCSO~J232612$-$531858 & 0.15 & 0.13 & $ 77.2\pm  8.8$ & $ 69.4\pm  8.3$ & $ 1.8 \times 10^{12}\pm3.2 \times 10^{11}$ & $2.5 \times 10^{14}$ & $3.1 \times 10^{14}$  \\
SCSO~J232627$-$531512 & 0.74 & 0.74 & $ 39.2\pm  6.3$ & $ 42.0\pm  6.5$ & $ 2.5 \times 10^{12}\pm2.3 \times 10^{11}$ & $3.2 \times 10^{14}$ & $6.6 \times 10^{14}$  \\
SCSO~J232637$-$533911 & 0.76 & 0.77 & $ 34.4\pm  5.9$ & $ 56.1\pm  7.5$ & $ 3.1 \times 10^{12}\pm2.2 \times 10^{11}$ & $4.5 \times 10^{14}$ & $8.6 \times 10^{14}$  \\
SCSO~J232653$-$524149 & 0.11 & 0.11 & $ 52.8\pm  7.3$ & $ 87.1\pm  9.3$ & $ 1.2 \times 10^{12}\pm1.8 \times 10^{11}$ & $3.1 \times 10^{14}$ & $2.2 \times 10^{14}$  \\
SCSO~J232724$-$533553 & 0.74 & 0.74 & $ 56.6\pm  7.5$ & $103.1\pm 10.2$ & $ 6.0 \times 10^{12}\pm5.4 \times 10^{11}$ & $9.2 \times 10^{14}$ & $2.0 \times 10^{15}$  \\
SCSO~J232800$-$535152 & 0.74 & 0.75 & $ 36.8\pm  6.1$ & $ 63.9\pm  8.0$ & $ 1.4 \times 10^{12}\pm2.0 \times 10^{11}$ & $3.7 \times 10^{14}$ & $2.8 \times 10^{14}$  \\
SCSO~J232811$-$533847 & 0.74 & 0.74 & $ 43.1\pm  6.6$ & $ 47.2\pm  6.9$ & $ 2.1 \times 10^{12}\pm2.3 \times 10^{11}$ & $3.4 \times 10^{14}$ & $5.2 \times 10^{14}$  \\
SCSO~J232825$-$522814 & 0.73 & 0.70 & $ 35.0\pm  5.9$ & $ 47.3\pm  6.9$ & $ 2.9 \times 10^{12}\pm1.4 \times 10^{11}$ & $4.3 \times 10^{14}$ & $9.0 \times 10^{14}$  \\
SCSO~J232827$-$531414 & 0.36 & 0.35 & $ 34.7\pm  5.9$ & $ 39.5\pm  6.3$ & $ 1.5 \times 10^{12}\pm1.1 \times 10^{11}$ & $3.0 \times 10^{14}$ & $3.7 \times 10^{14}$  \\
SCSO~J232856$-$552428 & 0.57 & 0.57 & $ 20.3\pm  4.5$ & $ 19.6\pm  4.4$ & $ 1.2 \times 10^{12}\pm5.2 \times 10^{10}$ & $2.0 \times 10^{14}$ & $3.5 \times 10^{14}$  \\
SCSO~J232916$-$522910 & 0.73 & 0.74 & $ 39.8\pm  6.3$ & $ 58.1\pm  7.6$ & $ 3.3 \times 10^{12}\pm1.6 \times 10^{11}$ & $4.5 \times 10^{14}$ & $9.2 \times 10^{14}$  \\
SCSO~J233006$-$524035 & 0.73 & 0.71 & $ 36.4\pm  6.0$ & $ 43.8\pm  6.6$ & $ 2.0 \times 10^{12}\pm1.4 \times 10^{11}$ & $3.2 \times 10^{14}$ & $5.0 \times 10^{14}$  \\
SCSO~J233227$-$535827 & 0.35 & 0.32 & $ 41.0\pm  6.4$ & $ 42.4\pm  6.5$ & $ 9.3 \times 10^{11}\pm9.0 \times 10^{10}$ & $3.7 \times 10^{14}$ & $2.4 \times 10^{14}$  \\
SCSO~J233232$-$522016 & 0.36 & 0.37 & $ 37.4\pm  6.1$ & $ 43.2\pm  6.6$ & $ 8.9 \times 10^{11}\pm8.2 \times 10^{10}$ & $4.0 \times 10^{14}$ & $2.4 \times 10^{14}$  \\
SCSO~J233329$-$521513 & 0.51 & 0.50 & $ 36.8\pm  6.1$ & $ 42.2\pm  6.5$ & $ 1.7 \times 10^{12}\pm1.0 \times 10^{11}$ & $3.6 \times 10^{14}$ & $4.6 \times 10^{14}$  \\
SCSO~J233420$-$542732 & 0.56 & 0.55 & $ 31.7\pm  5.6$ & $ 37.5\pm  6.1$ & $ 1.5 \times 10^{12}\pm9.8 \times 10^{10}$ & $2.7 \times 10^{14}$ & $3.6 \times 10^{14}$  \\
SCSO~J233544$-$535115 & 0.51 & 0.51 & $ 58.8\pm  7.7$ & $ 96.8\pm  9.8$ & $ 4.3 \times 10^{12}\pm2.1 \times 10^{11}$ & $8.1 \times 10^{14}$ & $1.3 \times 10^{15}$  \\
SCSO~J233556$-$560602 & 0.64 & 0.63 & $ 14.2\pm  3.8$ & $ 19.8\pm  4.5$ & $ 1.3 \times 10^{12}\pm7.3 \times 10^{10}$ & $2.5 \times 10^{14}$ & $4.7 \times 10^{14}$  \\
SCSO~J233607$-$535235 & 0.54 & 0.53 & $ 60.1\pm  7.8$ & $ 83.3\pm  9.1$ & $ 5.3 \times 10^{12}\pm2.6 \times 10^{11}$ & $1.0 \times 10^{15}$ & $2.2 \times 10^{15}$  \\
SCSO~J233726$-$565655 & 0.50 & 0.52 & $ 26.7\pm  5.2$ & $ 31.1\pm  5.6$ & $ 1.2 \times 10^{12}\pm7.6 \times 10^{10}$ & $2.7 \times 10^{14}$ & $3.0 \times 10^{14}$  \\
SCSO~J233807$-$531223 & 0.47 & 0.49 & $ 46.1\pm  6.8$ & $ 53.3\pm  7.3$ & $ 2.1 \times 10^{12}\pm1.8 \times 10^{11}$ & $4.8 \times 10^{14}$ & $6.2 \times 10^{14}$  \\
SCSO~J233931$-$544525 & 0.73 & 0.71 & $ 29.6\pm  5.4$ & $ 33.6\pm  5.8$ & $ 1.6 \times 10^{12}\pm1.1 \times 10^{11}$ & $2.7 \times 10^{14}$ & $4.1 \times 10^{14}$  \\
SCSO~J234012$-$541907 & 0.59 & 0.62 & $ 25.9\pm  5.1$ & $ 26.2\pm  5.1$ & $ 1.5 \times 10^{12}\pm9.8 \times 10^{10}$ & $2.6 \times 10^{14}$ & $4.4 \times 10^{14}$  \\
SCSO~J234138$-$545210 & 0.55 & 0.56 & $ 25.7\pm  5.1$ & $ 26.3\pm  5.1$ & $ 1.1 \times 10^{12}\pm7.8 \times 10^{10}$ & $2.8 \times 10^{14}$ & $3.5 \times 10^{14}$  \\
SCSO~J234156$-$530848 & 0.49 & 0.49 & $ 44.7\pm  6.7$ & $ 77.1\pm  8.8$ & $ 2.7 \times 10^{12}\pm1.7 \times 10^{11}$ & $7.8 \times 10^{14}$ & $8.9 \times 10^{14}$  \\
SCSO~J234703$-$535051 & 0.56 & 0.55 & $ 21.6\pm  4.7$ & $ 22.0\pm  4.7$ & $ 1.1 \times 10^{12}\pm5.6 \times 10^{10}$ & $2.0 \times 10^{14}$ & $3.0 \times 10^{14}$  \\
SCSO~J234917$-$545521 & 0.73 & 0.72 & $ 17.9\pm  4.2$ & $ 17.4\pm  4.2$ & $ 1.1 \times 10^{12}\pm6.7 \times 10^{10}$ & $1.8 \times 10^{14}$ & $3.2 \times 10^{14}$  \\
SCSO~J235055$-$530124 & 0.46 & 0.48 & $ 41.4\pm  6.4$ & $ 56.3\pm  7.5$ & $ 1.4 \times 10^{12}\pm1.1 \times 10^{11}$ & $4.8 \times 10^{14}$ & $3.6 \times 10^{14}$  \\
SCSO~J235138$-$545253 & 0.33 & 0.31 & $ 71.4\pm  8.4$ & $122.3\pm 11.1$ & $ 2.5 \times 10^{12}\pm2.3 \times 10^{11}$ & $1.1 \times 10^{15}$ & $7.4 \times 10^{14}$  \\
SCSO~J235233$-$564348 & 0.74 & 0.72 & $ 17.1\pm  4.1$ & $ 16.2\pm  4.0$ & $ 1.1 \times 10^{12}\pm5.2 \times 10^{10}$ & $1.7 \times 10^{14}$ & $3.1 \times 10^{14}$  \\
SCSO~J235454$-$563311 & 0.51 & 0.50 & $ 33.1\pm  5.8$ & $ 39.4\pm  6.3$ & $ 1.7 \times 10^{12}\pm8.0 \times 10^{10}$ & $3.8 \times 10^{14}$ & $5.0 \times 10^{14}$  \\
\enddata
\label{tab:clustes23hr}
\tablecomments{Catalog with the optical properties of clusters with
  mass estimates $>3\times 10^{14} M_{\sun}$ in the 23hr region. For
  each cluster we note the BCGs photometric redshift and the median
  photometric redshift for the clusters using the members within
  $250$~kpc of the center of the cluster. The ID is based on the
  position of the BCG.}
\end{deluxetable*}
%\end{deluxetable} % MS -- style

%%%%%%%%%%%%%%%%%%%%%%%%%%
%% 05hr clusters Table %%%
%%%%%%%%%%%%%%%%%%%%%%%%%%
\begin{deluxetable*}{rrrccrrrr}[b]
%\begin{deluxetable}{rrrccrrrr} % MS -- style
%\tabletypesize{\footnotesize } 
%\rotate
\tablecaption{Optical Clusters with $M(N_{200})>3\times
  10^{14}M_{\sun}$ or $M(L_{200})>3\times
  10^{14}M_{\sun}$ in the 5hr field}
\tablewidth{0pt}
\tablehead{
\colhead{ID} & 
\colhead{$z_{\rm cluster}$} & 
\colhead{$z_{\rm BCG}$} & 
\colhead{$N_{\rm gal}$} &
\colhead{\n200} &
\colhead{$L_{200}[L_{\sun}]$} &
\colhead{$M(N_{200})$} &
\colhead{$M(L_{200})$} \\
\multicolumn{6}{c}{} & \multicolumn{2}{c}{$(M_{\sun})$}  
}
\startdata
SCSO~J050854$-$513048 & 0.70 & 0.70 & $ 25.0\pm  5.0$ & $ 23.9\pm  4.9$ & $ 1.3 \times 10^{12}\pm1.4 \times 10^{11}$ & $2.4 \times 10^{14}$ & $3.8 \times 10^{14}$  \\
SCSO~J050857$-$535837 & 0.76 & 0.78 & $ 37.0\pm  6.1$ & $ 48.5\pm  7.0$ & $ 3.5 \times 10^{12}\pm2.4 \times 10^{11}$ & $5.0 \times 10^{14}$ & $1.2 \times 10^{15}$  \\
SCSO~J050902$-$520704 & 0.58 & 0.58 & $ 35.3\pm  5.9$ & $ 37.0\pm  6.1$ & $ 1.9 \times 10^{12}\pm1.7 \times 10^{11}$ & $2.8 \times 10^{14}$ & $4.8 \times 10^{14}$  \\
SCSO~J050926$-$522227 & 0.67 & 0.70 & $ 30.9\pm  5.6$ & $ 33.5\pm  5.8$ & $ 1.6 \times 10^{12}\pm1.8 \times 10^{11}$ & $2.5 \times 10^{14}$ & $4.0 \times 10^{14}$  \\
SCSO~J051023$-$544455 & 0.39 & 0.37 & $ 45.2\pm  6.7$ & $ 51.0\pm  7.1$ & $ 1.1 \times 10^{12}\pm1.7 \times 10^{11}$ & $3.5 \times 10^{14}$ & $2.3 \times 10^{14}$  \\
SCSO~J051112$-$523112 & 0.73 & 0.73 & $ 31.7\pm  5.6$ & $ 34.4\pm  5.9$ & $ 2.2 \times 10^{12}\pm2.2 \times 10^{11}$ & $3.3 \times 10^{14}$ & $6.6 \times 10^{14}$  \\
SCSO~J051136$-$561045 & 0.70 & 0.71 & $ 30.9\pm  5.6$ & $ 35.0\pm  5.9$ & $ 2.9 \times 10^{12}\pm1.6 \times 10^{11}$ & $3.2 \times 10^{14}$ & $8.7 \times 10^{14}$  \\
SCSO~J051144$-$511416 & 0.48 & 0.48 & $ 33.3\pm  5.8$ & $ 40.1\pm  6.3$ & $ 1.6 \times 10^{12}\pm1.8 \times 10^{11}$ & $2.9 \times 10^{14}$ & $3.7 \times 10^{14}$  \\
SCSO~J051145$-$515430 & 0.70 & 0.70 & $ 32.7\pm  5.7$ & $ 32.4\pm  5.7$ & $ 1.6 \times 10^{12}\pm2.1 \times 10^{11}$ & $2.2 \times 10^{14}$ & $3.5 \times 10^{14}$  \\
SCSO~J051207$-$514204 & 0.48 & 0.46 & $ 37.7\pm  6.1$ & $ 49.1\pm  7.0$ & $ 1.6 \times 10^{12}\pm1.9 \times 10^{11}$ & $3.5 \times 10^{14}$ & $3.9 \times 10^{14}$  \\
SCSO~J051225$-$505913 & 0.70 & 0.70 & $ 49.8\pm  7.0$ & $ 54.6\pm  7.4$ & $ 3.0 \times 10^{12}\pm3.2 \times 10^{11}$ & $4.9 \times 10^{14}$ & $9.2 \times 10^{14}$  \\
SCSO~J051240$-$513941 & 0.66 & 0.66 & $ 33.8\pm  5.8$ & $ 30.8\pm  5.5$ & $ 1.5 \times 10^{12}\pm1.9 \times 10^{11}$ & $2.2 \times 10^{14}$ & $3.5 \times 10^{14}$  \\
SCSO~J051245$-$502028 & 0.62 & 0.62 & $ 24.2\pm  4.9$ & $ 23.3\pm  4.8$ & $ 2.6 \times 10^{12}\pm5.4 \times 10^{10}$ & $2.8 \times 10^{14}$ & $9.4 \times 10^{14}$  \\
SCSO~J051258$-$542153 & 0.67 & 0.68 & $ 18.5\pm  4.3$ & $ 18.4\pm  4.3$ & $ 1.3 \times 10^{12}\pm1.2 \times 10^{11}$ & $2.0 \times 10^{14}$ & $4.0 \times 10^{14}$  \\
SCSO~J051400$-$512635 & 0.72 & 0.73 & $ 54.7\pm  7.4$ & $ 81.7\pm  9.0$ & $ 5.7 \times 10^{12}\pm4.9 \times 10^{11}$ & $7.0 \times 10^{14}$ & $1.8 \times 10^{15}$  \\
SCSO~J051412$-$514004 & 0.67 & 0.66 & $ 44.8\pm  6.7$ & $ 60.9\pm  7.8$ & $ 2.5 \times 10^{12}\pm3.1 \times 10^{11}$ & $4.2 \times 10^{14}$ & $6.0 \times 10^{14}$  \\
SCSO~J051457$-$514345 & 0.69 & 0.69 & $ 36.4\pm  6.0$ & $ 37.4\pm  6.1$ & $ 4.0 \times 10^{12}\pm3.4 \times 10^{11}$ & $4.1 \times 10^{14}$ & $1.5 \times 10^{15}$  \\
SCSO~J051542$-$514017 & 0.73 & 0.73 & $ 40.4\pm  6.3$ & $ 53.4\pm  7.3$ & $ 2.4 \times 10^{12}\pm3.3 \times 10^{11}$ & $4.0 \times 10^{14}$ & $6.3 \times 10^{14}$  \\
SCSO~J051558$-$543906 & 0.66 & 0.64 & $ 33.7\pm  5.8$ & $ 38.4\pm  6.2$ & $ 1.5 \times 10^{12}\pm1.8 \times 10^{11}$ & $2.8 \times 10^{14}$ & $3.5 \times 10^{14}$  \\
SCSO~J051613$-$542620 & 0.38 & 0.36 & $ 45.9\pm  6.8$ & $ 65.2\pm  8.1$ & $ 4.3 \times 10^{12}\pm2.4 \times 10^{11}$ & $6.8 \times 10^{14}$ & $1.5 \times 10^{15}$  \\
SCSO~J051637$-$543001 & 0.23 & 0.25 & $127.0\pm 11.3$ & $180.3\pm 13.4$ & $ 5.3 \times 10^{12}\pm1.0 \times 10^{12}$ & $1.8 \times 10^{15}$ & $1.9 \times 10^{15}$  \\
SCSO~J051755$-$555727 & 0.66 & 0.66 & $ 20.4\pm  4.5$ & $ 18.2\pm  4.3$ & $ 1.2 \times 10^{12}\pm1.1 \times 10^{11}$ & $1.7 \times 10^{14}$ & $3.4 \times 10^{14}$  \\
SCSO~J051933$-$554243 & 0.69 & 0.70 & $ 25.0\pm  5.0$ & $ 23.8\pm  4.9$ & $ 2.5 \times 10^{12}\pm1.8 \times 10^{11}$ & $2.9 \times 10^{14}$ & $9.1 \times 10^{14}$  \\
SCSO~J051935$-$554916 & 0.75 & 0.75 & $ 25.8\pm  5.1$ & $ 25.7\pm  5.1$ & $ 1.6 \times 10^{12}\pm1.7 \times 10^{11}$ & $2.2 \times 10^{14}$ & $4.3 \times 10^{14}$  \\
SCSO~J052051$-$561804 & 0.74 & 0.73 & $ 55.4\pm  7.5$ & $ 91.0\pm  9.5$ & $ 4.5 \times 10^{12}\pm3.7 \times 10^{11}$ & $7.1 \times 10^{14}$ & $1.3 \times 10^{15}$  \\
SCSO~J052113$-$510418 & 0.64 & 0.61 & $ 58.2\pm  7.6$ & $ 85.1\pm  9.2$ & $ 4.8 \times 10^{12}\pm4.9 \times 10^{11}$ & $7.9 \times 10^{14}$ & $1.6 \times 10^{15}$  \\
SCSO~J052200$-$502700 & 0.50 & 0.47 & $ 74.0\pm  8.6$ & $133.7\pm 11.6$ & $ 4.1 \times 10^{12}\pm4.0 \times 10^{11}$ & $9.4 \times 10^{14}$ & $1.1 \times 10^{15}$  \\
SCSO~J052533$-$551818 & 0.72 & 0.72 & $ 19.1\pm  4.4$ & $ 15.3\pm  3.9$ & $ 1.6 \times 10^{12}\pm1.4 \times 10^{11}$ & $1.5 \times 10^{14}$ & $4.8 \times 10^{14}$  \\
SCSO~J052608$-$561114 & 0.14 & 0.15 & $ 41.6\pm  6.5$ & $ 63.4\pm  8.0$ & $ 1.4 \times 10^{12}\pm2.6 \times 10^{11}$ & $3.5 \times 10^{14}$ & $3.2 \times 10^{14}$  \\
SCSO~J052803$-$525945 & 0.68 & 0.69 & $ 61.5\pm  7.8$ & $ 85.3\pm  9.2$ & $ 4.3 \times 10^{12}\pm4.4 \times 10^{11}$ & $6.8 \times 10^{14}$ & $1.2 \times 10^{15}$  \\
SCSO~J052810$-$514839 & 0.65 & 0.64 & $ 24.6\pm  5.0$ & $ 23.6\pm  4.8$ & $ 1.7 \times 10^{12}\pm1.3 \times 10^{11}$ & $2.3 \times 10^{14}$ & $5.0 \times 10^{14}$  \\
SCSO~J052858$-$535744 & 0.70 & 0.70 & $ 37.4\pm  6.1$ & $ 53.6\pm  7.3$ & $ 3.0 \times 10^{12}\pm3.1 \times 10^{11}$ & $4.2 \times 10^{14}$ & $8.1 \times 10^{14}$  \\
SCSO~J052951$-$551611 & 0.66 & 0.65 & $ 26.7\pm  5.2$ & $ 30.3\pm  5.5$ & $ 2.4 \times 10^{12}\pm1.9 \times 10^{11}$ & $3.3 \times 10^{14}$ & $8.1 \times 10^{14}$  \\
SCSO~J053052$-$552056 & 0.73 & 0.71 & $ 40.8\pm  6.4$ & $ 71.3\pm  8.4$ & $ 4.4 \times 10^{12}\pm4.2 \times 10^{11}$ & $5.8 \times 10^{14}$ & $1.3 \times 10^{15}$  \\
SCSO~J053154$-$552031 & 0.23 & 0.21 & $ 92.3\pm  9.6$ & $114.5\pm 10.7$ & $ 2.6 \times 10^{12}\pm5.1 \times 10^{11}$ & $7.9 \times 10^{14}$ & $6.6 \times 10^{14}$  \\
SCSO~J053327$-$542016 & 0.23 & 0.25 & $ 47.7\pm  6.9$ & $ 77.3\pm  8.8$ & $ 1.7 \times 10^{12}\pm3.3 \times 10^{11}$ & $3.3 \times 10^{14}$ & $3.4 \times 10^{14}$  \\
SCSO~J053437$-$552312 & 0.76 & 0.80 & $ 15.4\pm  3.9$ & $  8.4\pm  2.9$ & $ 1.4 \times 10^{12}\pm1.3 \times 10^{11}$ & $1.1 \times 10^{14}$ & $5.1 \times 10^{14}$  \\
SCSO~J053448$-$543534 & 0.65 & 0.65 & $ 23.0\pm  4.8$ & $ 24.0\pm  4.9$ & $ 1.3 \times 10^{12}\pm1.3 \times 10^{11}$ & $2.4 \times 10^{14}$ & $3.9 \times 10^{14}$  \\
SCSO~J053500$-$532018 & 0.59 & 0.57 & $ 23.1\pm  4.8$ & $ 22.8\pm  4.8$ & $ 3.2 \times 10^{12}\pm1.3 \times 10^{11}$ & $2.9 \times 10^{14}$ & $1.2 \times 10^{15}$  \\
SCSO~J053632$-$553123 & 0.72 & 0.72 & $ 54.5\pm  7.4$ & $ 98.8\pm  9.9$ & $ 4.5 \times 10^{12}\pm5.7 \times 10^{11}$ & $6.2 \times 10^{14}$ & $1.1 \times 10^{15}$  \\
SCSO~J053638$-$553854 & 0.74 & 0.71 & $ 40.7\pm  6.4$ & $ 65.9\pm  8.1$ & $ 2.2 \times 10^{12}\pm3.1 \times 10^{11}$ & $3.9 \times 10^{14}$ & $4.8 \times 10^{14}$  \\
SCSO~J053645$-$553302 & 0.74 & 0.71 & $ 37.1\pm  6.1$ & $ 62.7\pm  7.9$ & $ 2.0 \times 10^{12}\pm3.2 \times 10^{11}$ & $4.4 \times 10^{14}$ & $4.9 \times 10^{14}$  \\
SCSO~J053655$-$553809 & 0.76 & 0.72 & $ 41.8\pm  6.5$ & $ 64.7\pm  8.1$ & $ 2.9 \times 10^{12}\pm3.2 \times 10^{11}$ & $4.6 \times 10^{14}$ & $7.2 \times 10^{14}$  \\
SCSO~J053715$-$541530 & 0.49 & 0.51 & $ 21.1\pm  4.6$ & $ 20.1\pm  4.5$ & $ 1.1 \times 10^{12}\pm1.1 \times 10^{11}$ & $2.0 \times 10^{14}$ & $3.1 \times 10^{14}$  \\
SCSO~J053732$-$542521 & 0.62 & 0.61 & $ 21.5\pm  4.6$ & $ 21.4\pm  4.6$ & $ 2.4 \times 10^{12}\pm1.3 \times 10^{11}$ & $2.5 \times 10^{14}$ & $8.6 \times 10^{14}$  \\
SCSO~J053952$-$561423 & 0.36 & 0.36 & $ 36.9\pm  6.1$ & $ 48.4\pm  7.0$ & $ 1.1 \times 10^{12}\pm1.4 \times 10^{11}$ & $4.1 \times 10^{14}$ & $2.9 \times 10^{14}$  \\
SCSO~J054012$-$561700 & 0.38 & 0.38 & $ 51.4\pm  7.2$ & $ 59.4\pm  7.7$ & $ 1.5 \times 10^{12}\pm1.8 \times 10^{11}$ & $5.0 \times 10^{14}$ & $3.9 \times 10^{14}$  \\
SCSO~J054022$-$541622 & 0.51 & 0.48 & $ 35.1\pm  5.9$ & $ 46.1\pm  6.8$ & $ 1.8 \times 10^{12}\pm1.9 \times 10^{11}$ & $3.3 \times 10^{14}$ & $4.3 \times 10^{14}$  \\
SCSO~J054052$-$551943 & 0.76 & 0.78 & $ 40.9\pm  6.4$ & $ 57.5\pm  7.6$ & $ 3.0 \times 10^{12}\pm3.2 \times 10^{11}$ & $4.5 \times 10^{14}$ & $8.1 \times 10^{14}$  \\
SCSO~J054228$-$525002 & 0.65 & 0.66 & $ 21.8\pm  4.7$ & $ 21.3\pm  4.6$ & $ 1.7 \times 10^{12}\pm1.3 \times 10^{11}$ & $1.8 \times 10^{14}$ & $4.5 \times 10^{14}$  \\
SCSO~J054332$-$505651 & 0.35 & 0.36 & $ 45.5\pm  6.8$ & $ 64.0\pm  8.0$ & $ 1.3 \times 10^{12}\pm1.9 \times 10^{11}$ & $4.1 \times 10^{14}$ & $2.8 \times 10^{14}$  \\
SCSO~J054358$-$531349 & 0.24 & 0.25 & $ 30.0\pm  5.5$ & $ 41.2\pm  6.4$ & $ 9.9 \times 10^{11}\pm1.8 \times 10^{11}$ & $3.1 \times 10^{14}$ & $2.3 \times 10^{14}$  \\
SCSO~J054401$-$511254 & 0.28 & 0.28 & $ 31.4\pm  5.6$ & $ 43.3\pm  6.6$ & $ 1.2 \times 10^{12}\pm1.6 \times 10^{11}$ & $3.4 \times 10^{14}$ & $3.0 \times 10^{14}$  \\
SCSO~J054407$-$530924 & 0.25 & 0.26 & $ 42.9\pm  6.5$ & $ 47.4\pm  6.9$ & $ 1.0 \times 10^{12}\pm1.9 \times 10^{11}$ & $3.3 \times 10^{14}$ & $2.2 \times 10^{14}$  \\
SCSO~J054436$-$550319 & 0.35 & 0.36 & $ 33.0\pm  5.8$ & $ 41.5\pm  6.4$ & $ 1.9 \times 10^{12}\pm1.6 \times 10^{11}$ & $3.6 \times 10^{14}$ & $5.3 \times 10^{14}$  \\
SCSO~J054721$-$554906 & 0.59 & 0.59 & $ 38.1\pm  6.2$ & $ 49.2\pm  7.0$ & $ 2.4 \times 10^{12}\pm2.1 \times 10^{11}$ & $4.2 \times 10^{14}$ & $6.8 \times 10^{14}$  \\
SCSO~J054742$-$554836 & 0.50 & 0.50 & $ 26.6\pm  5.2$ & $ 25.8\pm  5.1$ & $ 1.3 \times 10^{12}\pm1.2 \times 10^{11}$ & $2.5 \times 10^{14}$ & $3.9 \times 10^{14}$  \\
SCSO~J054811$-$555601 & 0.64 & 0.64 & $ 27.4\pm  5.2$ & $ 28.7\pm  5.3$ & $ 2.2 \times 10^{12}\pm1.5 \times 10^{11}$ & $2.8 \times 10^{14}$ & $6.6 \times 10^{14}$  \\
SCSO~J054931$-$522655 & 0.38 & 0.39 & $ 21.2\pm  4.6$ & $ 22.0\pm  4.7$ & $ 2.1 \times 10^{12}\pm1.1 \times 10^{11}$ & $2.7 \times 10^{14}$ & $7.5 \times 10^{14}$  \\
SCSO~J054949$-$513503 & 0.28 & 0.28 & $ 38.4\pm  6.2$ & $ 51.4\pm  7.2$ & $ 1.3 \times 10^{12}\pm1.8 \times 10^{11}$ & $4.2 \times 10^{14}$ & $3.3 \times 10^{14}$  \\
SCSO~J055017$-$534601 & 0.49 & 0.47 & $ 30.8\pm  5.5$ & $ 29.9\pm  5.5$ & $ 1.4 \times 10^{12}\pm1.4 \times 10^{11}$ & $2.5 \times 10^{14}$ & $3.6 \times 10^{14}$  \\
\enddata
\label{tab:clustes05hr}
\tablecomments{Catalog with the optical properties of clusters with
  mass estimates $>3\times 10^{14} M_{\sun}$ in the 5hr region. For
  each cluster we note the BCGs photometric redshift and the median
  photometric redshift for the clusters using the members within
  $250$~kpc of the center of the cluster. The ID is based on the
  position of the BCG.}
%\end{deluxetable}
\end{deluxetable*}

\subsection{Correlation with Known Sources}

We queried the NASA/IPCA Extragalactic Database
(NED)\footnote{http://nedwww.ipac.caltech.edu/} for catalogued
clusters from {\em ROSAT}, \citet{ACO89}, ACT \citep{Hincks09} and SPT
\citep{Stan09,MenanteauHughes09} within a 3$^\prime$ radius of the
location of each SCS cluster (see Table~\ref{tab:clustersID}). In some
cases there was a catalogued galaxy from the 2 Micron All Sky Survey
\citep[][NED ID: 2MASX]{2MASS}, which we report if it is within $10''$
of the BCG.
We also take note of radio sources within $1'$ of our cluster
positions since these could potentially bias the cluster SZ signal. We
only found radio sources from the Sydney University Molonglo Sky
Survey (SUMSS) at 843~MHz \citep{SUMSS03}. Finally we find one
unidentified bright X-ray source from the {\em ROSAT} All Sky Survey
(RASS) \citep{Voges99} that is coincident with one of our clusters.

It is interesting to note that although we recover two of the
SZE-selected clusters from the first SPT results \citep{Stan09}, there
are two others from that study which we do not recover.  One of these
(SPT-CL 0509$-$5342) falls below the mass threshold used here with
mass estimates of $2.1\times10^{14}M_{\sun}$ and
$1.2\times10^{14}M_{\sun}$ for \MN200 and \ML200 respectively. These
are lower than the optical mass quoted in \cite{MenanteauHughes09} due
to the change in the \cite{Reyes08} mass parametrization but still are
comparable to the weak lensing mass range,
$1.9-5.6\times10^{14}~M_{\sun}$, obtained by \cite{McInnes09} using
the same optical dataset.
The other cluster (SPT-CL0547$-$5345) has a photometric redshift of
0.88 \citep{MenanteauHughes09} that puts it beyond the redshift
threshold ($z=0.8$) we use here.

\subsection{ROSAT archival data}

\begin{figure}
%% \centerline{\includegraphics[width=5.0in]{f7.eps}}
\centerline{\includegraphics[width=3.5in]{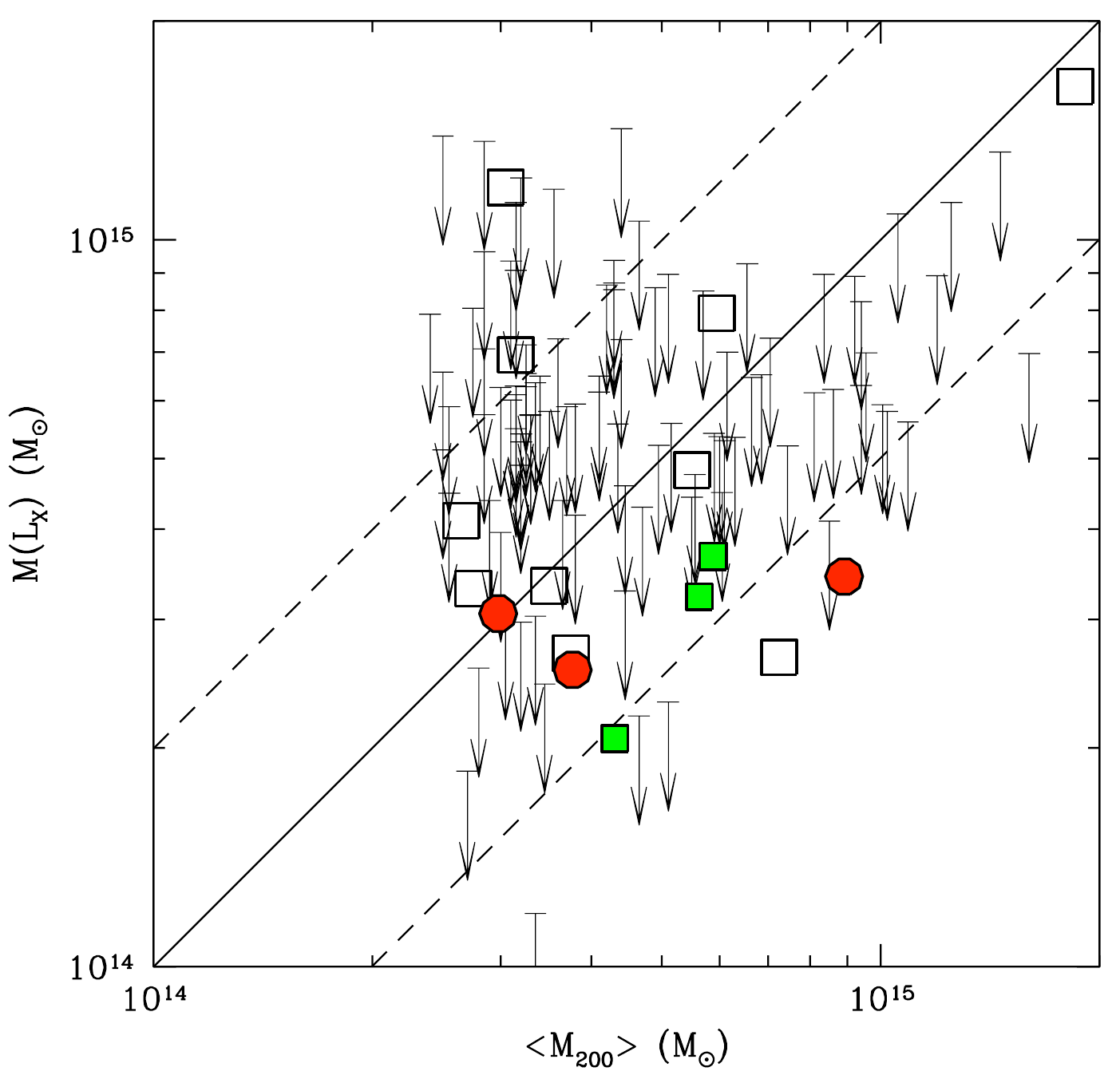}}
\caption{Plot of X-ray-- vs.~optically--derived masses for all SCS
  clusters.  The optical mass is the average of \MN200 and \ML200,
  while the X-ray mass comes from the X-ray luminosity assuming a
  $M$-$L_X$ scaling relation.  Clusters with X-ray emission are shown
  as the box symbols, while the others are shown as upper limits. The
  solid line denotes equality; the dashed lines indicate a factor of
  two range in mass. The green points show the average optical and
  X-ray masses for clusters stacked in three redshift bins (0.0-0.4,
  0.4-0.6, 0.6-0.8) ordered from bottom to top, while the red points
  were stacked in three optical mass bins ($<$3.2$\times 10^{14}
  M_{\sun}$, 3.2$\times 10^{14} M_{\sun}$-4.5$\times 10^{14}
  M_{\sun}$, and $>$4.5$\times 10^{14} M_{\sun}$) ordered from left to
  right. }
\label{fig:masses}
\end{figure}

We searched for X-ray counterparts to the SCS optical clusters using
the {\em ROSAT} All Sky Survey (RASS) data following the same
procedure as in \citet{MenanteauHughes09}.  The raw X-ray photon event
lists and exposure maps were downloaded from the MPE {\em ROSAT}
Archive\footnote{ftp://ftp.xray.mpe.mpg.de/rosat/archive/} and queried
with our own custom software.  At the position of each SCS cluster,
RASS count rates in the 0.5--2 keV band (corresponding to PI channels
52--201) were extracted from within a 3$^\prime$ radius for the source
emission and from within a surrounding annulus (5$^\prime$ to
25$^\prime$ inner and outer radii) for the background emission.  The
background-subtracted count rates were converted to X-ray luminosity
(in the 0.5--2.0 keV band) assuming a thermal spectrum ($kT=5$ keV)
and the Galactic column density of neutral hydrogen appropriate to the
source position, using data from the Leiden/Argentine/ HI Bonn survey
\citep{kalberla05}.  X-ray masses within an overdensity of 500 times
the critical density were then determined using the $L_X$
vs.~$M_{500}$ scaling relation from \citet{Vikhlinin09}.  These were
then converted to an overdensity of 200 with respect to the average
density of the Universe for comparison to the optically-derived masses
using a simple multiplicative scaling factor of 1.77.  This factor is
good to 10\% over the redshift and mass range of our sample.  In
Table~\ref{tab:clustersRASS} we show redshifts, column densities,
rates, luminosities, and masses only for those clusters with X-ray
emission significant at 2 $\sigma$ or higher, while
Figure~\ref{fig:masses} shows all the SCS clusters with either actual
values for the X-ray mass or (for most cases) upper limits. We also
show points corresponding to a simple stacking of the X-ray and the
optical masses for clusters in three redshift and three optical mass
bins as the green and red dots, respectively.  Each point represents
an average of $\sim$20 clusters each with some positive X-ray count
rate, excluding the several higher significance individual cases
plotted in the Figure.  It is encouraging that the points scatter by
about a factor of two around the line of equal optical and X-ray
masses, suggesting that the X-ray emission is in fact associated with
hot gas in the potential well of a massive cluster.  In at least two
cases, we have recovered known clusters (SCSO J051637$-$543001 and
SCSO J232653$-$524149).  And two of the RASS-correlated clusters
(SCSO~J053154$-$552031 and SCSO~J233227$-$535827) are among those
shown in the color images (i.e., see the bottom left panel of
Fig.\ref{fig:clusters05hr} and the top panel of
Fig.~\ref{fig:clusters23hr}).

% M_X vs M_opt

\subsection{XMM-Newton archival data}

There are a number of {\em XMM-Newton} pointings that overlap with the
SCS especially in the 23-hr region.  Table~\ref{tab:clustersXMM} lists
all the optical clusters from this study (Tables~\ref{tab:clustes23hr}
and \ref{tab:clustes05hr}) that are located within the field of view
of an archival {\em XMM-Newton} pointing (specifically the following
ObsIds: 0205330301, 0505380601, 0505381801, 0505382201, and
0505383601).  In several cases, there is an associated X-ray source
detected by the Pipeline Processing System (PPS) which is run on all
{\em XMM-Newton} pointings to produce standard results and are
provided to the observer.  If a PPS source exists, we list its ID, the
offset in arcseconds between the X-ray source and the optical cluster
position and whether or not the PPS has flagged the X-ray source are
being extended.  To determine fluxes and luminosities, we extract
spectra from circular regions chosen to maximize the extracted count
rate for the extended X-ray sources.  For the other sources we use a
1$^\prime$ radius circle.  Background spectra come from a surrounding
annular region with sufficient area to obtain good photon statistics.
The rate, summed over the single PN and both MOS detectors, is quoted in
column 7 of Table~\ref{tab:clustersXMM}.  We also used standard {\em
  XMM-Newton} software tools to calculate the instrumental response
functions (i.e., the arf and rmf files) for each cluster.  X-ray
luminosity values were determined using the extracted spectra and
response files assuming a simple source emission model consisting of
an absorbed thermal plasma model. % (i.e., mekal). 
For the absorption component we fixed the column density of neutral
hydrogen to the Galactic value toward the cluster position (obtained
in the same way as for the RASS discussed above).  Three clusters have
enough signal that their mean temperatures can be measured (see
below); for the others we assume $kT=5$ keV for the luminosity
calculation.  The final column lists the cluster mass inferred from
the X-ray luminosity as discussed above.

The {\em XMM-Newton} spectra constrain the cluster temperatures for
\mbox{SCSO~J051558$-$543906} and \mbox{SCSO~J231651$-$545356} to
values of $kT = 1.8^{+0.5}_{-0.3}$ keV and $kT = 3.7^{+0.6}_{-0.5}$
keV, respectively (errors are at the 68\% confidence interval), assuming a
metal abundance of 0.3 times the solar value.  SCSO~J051637$-$543001
(Abell S0520) is a bright cluster that was the target of a specific
{\em XMM-Newton} program.  Our analysis finds a best fit temperature
of $kT = 7.7\pm0.3$ keV and metal abundance of $0.17 \pm 0.04$
relative to solar, which are both consistent with previous work
\citep[e.g.,][]{Zhang06}.  The X-ray masses for these three systems
just mentioned agree quite well with the optically-derived masses and,
in the case of SCSO~J051637$-$543001, with the ROSAT-derived mass as well.
For the three other clusters in the 23hr region in
Table~\ref{tab:clustersXMM} the X-ray masses are all
lower than the optical masses by factors of 3--5.  For the remaining
system (SCSO~J051613$-$542620) there is a catastrophic disagreement
between the inferred optical and X-ray masses with the X-ray mass
being more than an order of magnitude below the optical one.  This is
likely due to significant contamination of the optical number counts
by cluster members from the nearby, rich system Abell S0520 which is
also close in redshift (0.2952 vs.~0.36).

\section{Summary}

We have fully processed using an independent custom-built pipeline
$\sim1$ TB of archival {\em griz} imaging data from the CTIO Blanco
4-m telescope acquired under the NOAO Large Survey program (05B-0043,
PI: Joe Mohr), as part of our own Southern Cosmology Survey.  This
data volume corresponds to 45 nights of observing over 3 years
(2005-2007) and covers 70 deg$^2$ of the southern sky that has
also been fully observed by ACT and SPT.  Here we have presented the
first results from the full nominal data set, namely a sample of 105
massive, optically-selected galaxy clusters.  Future studies will
present the properties of lower mass clusters and groups as well as
multi-wavelength studies of cluster physics utilizing selected
clusters from this sample.

The current sample is limited to systems with optically-derived masses
greater than $3\times 10^{14}\, M_\sun$ and redshifts less than $0.8$.
We have chosen this mass limit to be at or below the anticipated mass
threshold of ACT and SPT in order to encompass the upcoming
significant SZE detections. However we also have a redshift limit
which is due to the depth of the imaging and the wavelength coverage
of the filter set. Thus we are missing the most interesting massive
clusters at high redshift ($z>0.8$).  However, as demonstrated in
\citet{MenanteauHughes09}, the optical data analyzed here can be used
to confirm the presence of a cluster when conducting a targeted
positional search for a high significance SZE candidate.

The recent success of the mm-band wide-survey-area experiments (ACT
and SPT) in finding new clusters through untargeted SZE surveys has
been a strong catalyst for our work.  We present this cluster sample
to aid in the verification of SZE cluster candidates and the
characterization of the SZE selection function which currently is
observationally unexplored. Moreover, we anticipate stacking the ACT
mm-band maps at the positions of optical clusters to detect,
statistically, the average SZE signal for systems that fall below the
ACT detection threshold for individual sources.

Given the large volume of this data set we believe it might be helpful
to address other problems in astrophysics; therefore we plan to make
the Blanco data products (i.e., photometric source catalogs and
images) available to the community in December 2010
at the following URL http://scs.rutgers.edu.

\acknowledgments

We would like to thank the Blanco Cosmology Survey team for the
planning and execution of the CTIO Blanco observations that were used
in this paper.
We have made use of the ROSAT Data Archive of the Max-Planck-Institut
f{\"u}r extraterrestrische Physik (MPE) at Garching, Germany as well as
results obtained from the High Energy Astrophysics Science Archive
Research Center, provided by NASA's Goddard Space Flight
Center.
% NVO
This research has made use of data obtained from or software provided
by the US National Virtual Observatory, which is sponsored by the
National Science Foundation.
% NED
This research has made use of the  NASA/IPAC Extragalactic Database
(NED) which is operated by the Jet Propulsion Laboratory, California
Institute of Technology, under contract with the National Aeronautics
and Space Administration.
Partial financial support was provided by the National Science Foundation
under the PIRE program (award number OISE-0530095). We also
acknowledge support from NASA/XMM grants NNX08AX55G and NNX08AX72G.

%%%%%%%%%%%%%%%%%%%%%%%%%%%%%%%%%%%%%%%%%%%%%%%%%%%%%%%%%%%%%%
%% NED sources correlated with SCSO clusters  (within ~1') %%%
%% Excludes 2MASS and APMUKS sources                       %%%
%%%%%%%%%%%%%%%%%%%%%%%%%%%%%%%%%%%%%%%%%%%%%%%%%%%%%%%%%%%%%%
\begin{deluxetable*}{rllc}[b!]
%\begin{deluxetable}{rllc}
%\tabletypesize{\footnotesize } 
%\rotate
\tablecaption{Catalogued Sources Associated with SCS Optical Clusters}
\tablewidth{0pt}
\tablehead{
\colhead{SCS ID} & 
\colhead{Catalog name} &
\colhead{Distance to BCG} &
\colhead{Source type} 
}
\startdata
SCSO~J051412$-$514004 & SUMSS J051411$-$513953    & 13$^{\prime\prime}$ & RadioS  \\
SCSO~J051637$-$543001 & 2MASX J05163736$-$5430017 & 3$^{\prime\prime}$ & Galaxy \\
 \multicolumn{1}{c}{$^{\prime\prime}$}  & Abell S0520, RXC J0516.6$-$5430 &2.4$^{\prime}$ & GClstr ($z=0.2952$) \\
 \multicolumn{1}{c}{$^{\prime\prime}$}  & ACT-CL J0516$-$5432,SPT-CL 0517$-$5430  & 2.8$^\prime$,0.5$^\prime$ & SZ-GClstr \\
% \multicolumn{1}{c}{$^{\prime\prime}$}  & SPT-CL 0517$-$5430 & SZ-GClstr \\
SCSO~J052113$-$510418 & SUMSS J052114$-$510419 &13$^{\prime\prime}$ & RadioS  \\
SCSO~J052533$-$551818 & Abell S0529           &1.5$^{\prime}$  & GClstr  \\
SCSO~J052608$-$561114 & Abell S0530           &2.4$^{\prime}$  & GClstr  \\
SCSO~J052803$-$525945 & SUMSS J052805$-$525953  &24$^{\prime\prime}$ & RadioS  \\
 \multicolumn{1}{c}{$^{\prime\prime}$}  & SPT-CL 0528$-$5300 & 0.4$^\prime$  & SZ-GClstr  \\
SCSO~J053327$-$542016 & 2MASX J05332723$-$5420154  &2$^{\prime\prime}$ & Galaxy \\
SCSO~J053437$-$552312 & SUMSS J053442$-$552248  &55$^{\prime\prime}$ & RadioS  \\
SCSO~J053632$-$553123 & SUMSS J053629$-$553147  &30$^{\prime\prime}$ & RadioS  \\
SCSO~J053715$-$541530 & SUMSS J053718$-$541608  &48$^{\prime\prime}$ & RadioS  \\
SCSO~J054012$-$561700 & SUMSS J054014$-$561723  &31$^{\prime\prime}$ & RadioS  \\
SCSO~J054407$-$530924 & SUMSS J054406$-$530922  &8$^{\prime\prime}$ & RadioS  \\
SCSO~J054949$-$513503 & SUMSS J054948$-$513454  &9$^{\prime\prime}$ & RadioS  \\
SCSO~J055017$-$534601 & SUMSS J055019$-$534601  &25$^{\prime\prime}$ & RadioS  \\
SCSO~J231455$-$555308 & 2MASX J23145553$-$5553093   &5$^{\prime\prime}$ & Galaxy  \\
SCSO~J232001$-$565222 & SUMSS J232001$-$565219  &4$^{\prime\prime}$ & RadioS  \\
SCSO~J232653$-$524149 & RXC J2326.7$-$5242  &1.2$^{\prime}$   & GClstr ($z=0.1074$) \\
 \multicolumn{1}{c}{$^{\prime\prime}$}  & SUMSS J232654$-$524153  &14$^{\prime\prime}$ & RadioS \\
SCSO~J233227$-$535827 & 1RXS J233224.3$-$535840 &27$^{\prime\prime}$ & XrayS    \\
SCSO~J233544$-$535115 & SUMSS J233544$-$535113  &3$^{\prime\prime}$ & RadioS  \\
SCSO~J234156$-$530848 & SUMSS J234156$-$530849  &5$^{\prime\prime}$ & RadioS  \\
SCSO~J234703$-$535051 & SUMSS J234703$-$535052  &3$^{\prime\prime}$ & RadioS  \\
SCSO~J234917$-$545521 & SUMSS J234917$-$545518  &8$^{\prime\prime}$ & RadioS  \\
SCSO~J235055$-$530124 & SUMSS J235054$-$530141  &17$^{\prime\prime}$ & RadioS  \\
SCSO~J235138$-$545253 & 2MASX J23513813$-$5452538 &2$^{\prime\prime}$ & Galaxy  \\
\multicolumn{1}{c}{$^{\prime\prime}$} & SUMSS J235138$-$545255  &3$^{\prime\prime}$ & RadioS  \\

\enddata
\label{tab:clustersID}
\tablecomments{Catalogued sources from the NASA/IPAC Extragalactic Database (NED) correlated with 
  SCS optical clusters.}
\end{deluxetable*}
%\end{deluxetable}

%%%%%%%%%%%%%%%%%%%%%%%%%%
%% RASS X-ray clusters Table %%%
%%%%%%%%%%%%%%%%%%%%%%%%%%
\begin{deluxetable*}{rccccr}[b!]
%\begin{deluxetable}{rccccr}
%\tabletypesize{\footnotesize } 
%\rotate
\tablecaption{Optical Clusters with X-ray counterparts from the ROSAT All Sky Survey}
\tablewidth{0pt}
\tablehead{
\colhead{SCS ID} & 
\colhead{$z_{\rm cluster}$} & 
\colhead{$N_{\rm H}$} &
\colhead{Rate} &
\colhead{$L_{X}$(0.5--2.0 keV)} &
\colhead{$M_{200}(L_{X})$} \\
 & & ($10^{20}\,\rm cm^{-2}$) & (cts s$^{-1}$) & ($10^{44}\,\rm erg\, s^{-1}$) & $(M_{\sun})$
}
\startdata
SCSO~J051136$-$561045 &  0.70       & 1.61 & $0.015 \pm  0.006$ & $3.0 \pm 1.3$  &  $7.9\times 10^{14}$ \\
SCSO~J051637$-$543001 &  0.2952$^*$ & 2.05 & $0.205 \pm  0.030$ & $6.1 \pm 0.9$  &  $1.6\times 10^{15}$ \\
SCSO~J052533$-$551818 &  0.72       & 4.08 & $0.011 \pm  0.005$ & $2.5 \pm 1.2$  &  $6.9\times 10^{14}$ \\
SCSO~J053154$-$552031 &  0.23       & 5.20 & $0.017 \pm  0.006$ & $0.3 \pm 0.1$  &  $2.7\times 10^{14}$ \\
SCSO~J053952$-$561423 &  0.36       & 4.90 & $0.010 \pm  0.005$ & $0.5 \pm 0.3$  &  $3.3\times 10^{14}$ \\
SCSO~J054407$-$530924 &  0.25       & 5.01 & $0.020 \pm  0.005$ & $0.5 \pm 0.1$  &  $3.3\times 10^{14}$ \\
SCSO~J054721$-$554906 &  0.59       & 6.98 & $0.007 \pm  0.004$ & $1.2 \pm 0.6$  &  $4.8\times 10^{14}$ \\
SCSO~J054949$-$513503 &  0.28       & 4.58 & $0.012 \pm  0.004$ & $0.3 \pm 0.1$  &  $2.7\times 10^{14}$ \\
SCSO~J232653$-$524149 &  0.1074$^*$ & 1.28 & $0.166 \pm  0.035$ & $0.6 \pm 0.1$  &  $4.1\times 10^{14}$ \\
SCSO~J233227$-$535827 &  0.35       & 1.28 & $0.091 \pm  0.026$ & $3.9 \pm 1.1$  &  $1.2\times 10^{15}$ \\
\enddata
\label{tab:clustersRASS}
\tablecomments{Catalog of the optical clusters associated with X-ray
  emission in the ROSAT All Sky Survey.  The quoted redshifts are just
  repeated from Tables~\ref{tab:clustes23hr} and \ref{tab:clustes05hr}
  except for the starred values which are spectroscopic redshifts from
  NED (see Table~\ref{tab:clustersID}). Rates were extracted from
  within $3^\prime$ radii circles except for the three brightest
  clusters, SCSO~J051637$-$543001, SCSO~J232653$-$524149, and
  SCSO~J233227$-$535827, for which radii of $10^\prime$, $7^\prime$,
  and $6^\prime$, respectively, were used.  }
\end{deluxetable*}
%\end{deluxetable}

%%%%%%%%%%%%%%%%%%%%%%%%%%%%%%%%%%%%%%%%%%%%%%%%%%%%%%%%%%%%%
% XMM counterparts table from Amruta
%%%%%%%%%%%%%%%%%%%%%%%%%%%%%%%%%%%%%%%%%%%%%%%%%%%%%%%%%%%%%
\begin{turnpage} % comment in ms mode
\begin{deluxetable*}{ccccccrccccc}
%\begin{deluxetable}{cccccrcccccc}
%\tabletypesize{\footnotesize } 
%\rotate
\tablecaption{Optical Clusters Located within the Field of View of XMM-Newton Observations}
\tablewidth{0pt}
\tablehead{
\colhead{SCS ID} & 
\colhead{$z_{\rm cluster}$} & 
\colhead{$N_{\rm H}$} & 
\colhead{X-ray ID} & 
\colhead{Offset} & 
\colhead{Extended} & 
\colhead{Extr Rad} & 
\colhead{Rate (0.5--2 keV)} & 
\colhead{$L_{X}$ (0.5--2.0 keV) } &
\colhead{$M(L_{X})$} \\
% Second line
\colhead{} & 
\colhead{} & 
\colhead{($10^{20}$ cm$^{-2}$)} & 
\colhead{} & 
\colhead{($''$)} & 
\colhead{} & 
\colhead{(kpc)} & 
\colhead{($10^{-3}$cts s$^{-1}$)} & 
\colhead{($10^{44}$ ergs s$^{-1}$)} &
\colhead{($10^{14}\, M_{\sun}$)} \\
}
\startdata
  SCSO~J051558$-$543906 & $0.64$     & $2.07$ & XMM J051600$-$543900 & $19$     &  yes      &  910 (2.2$^\prime$) &  $15.7 \pm 2.6$ & $0.60$  & 3.0 \\%& \nodata \\
  SCSO~J051613$-$542620 & $0.36$     & $1.98$ & \nodata              & \nodata  &  \nodata  &  300 (1$^\prime$)   &   $6.8 \pm 1.4$ & $0.027$ & 0.5 \\%& \nodata \\
  SCSO~J051637$-$543001 & $0.2952^*$ & $2.05$ & XMM J051635$-$543022 & $25$     &  yes      & 1390 (5.3$^\prime$) &  $1740 \pm 15$  & $5.1$   & 14.4 \\%& \nodata \\
  SCSO~J231651$-$545356 & $0.36$     & $1.29$ & XMM J231653$-$545410 & $26$     &  yes      &  650 (2.2$^\prime$) &  $73.7 \pm 3.5$ & $0.71$  & 4.1 \\%& \nodata \\
  SCSO~J232856$-$552428 & $0.57$     & $1.29$ & XMM J232856$-$552429 & $ 5$     &  no       &  390 (1$^\prime$)   &   $5.4 \pm 1.3$ & $0.07$  & 0.8 \\%& \nodata \\
  SCSO~J233420$-$542732 & $0.55$     & $1.27$ & \nodata              & \nodata  &  \nodata  &  380 (1$^\prime$)   &   $5.2 \pm 1.4$ & $0.08$  & 0.9 \\%& \nodata \\
  SCSO~J233556$-$560602 & $0.63$     & $1.27$ & \nodata              & \nodata  &  \nodata  &  410 (1$^\prime$)   &   $6.2 \pm 1.7$ & $0.10$  & 1.0 \\%& \nodata \\
\enddata
\label{tab:clustersXMM}
\tablecomments{Luminosities calculated by fitting an absorbed thermal
  emission model with a fixed 5 keV temperature, except for SCSO~J051558$-$543906,
  SCSO~J051637$-$543001, and SCSO~J231651$-$545356 for which there was
  enough signal to determine their temperatures (see text).  The
  quoted redshifts are just repeated from Tables~\ref{tab:clustes23hr}
  and \ref{tab:clustes05hr} except for the starred value which is a
  spectroscopic redshift from NED (see Table~\ref{tab:clustersID}). }
\end{deluxetable*}
%\end{deluxetable}
\end{turnpage} % comment in ms mode

\end{document}